\documentclass[sn-mathphys-num]{sn-jnl}

\usepackage{graphicx}%
\usepackage{multirow}%
\usepackage{amsmath,amssymb,amsfonts}%
\usepackage{amsthm}%
\usepackage{mathrsfs}%
\usepackage[title]{appendix}%
\usepackage{xcolor}%
\usepackage{textcomp}%
\usepackage{manyfoot}%
\usepackage{booktabs}%
\usepackage{algorithm}%
\usepackage{algorithmicx}%
\usepackage{algpseudocode}%
\usepackage{listings}%

\def\bs{\boldsymbol}

\theoremstyle{thmstyleone}%
\theoremstyle{thmstyletwo}%

\theoremstyle{thmstylethree}%
\newtheorem{definition}{Definition}%

\begin{document}

\title[Geodesics' Gramian Denoising]{Image Denoising Using the Geodesics' Gramian of the Manifold Underlying Patch-Space}


\author*[1]{\fnm{Kelum} \sur{Gajamannage}}\email{kelum.gajamannage@uri.edu}

\affil*[1]{\orgdiv{Department of Mathematics and Applied Mathematical Sciences}, \orgname{University of Rhode Island}, \orgaddress{\street{45 Upper College Rd}, \city{Kingston}, \postcode{02881}, \state{RI}, \country{USA}}}


\abstract{With the proliferation of sophisticated cameras in modern society, the demand for accurate and visually pleasing images is increasing. However, the quality of an image captured by a camera may be degraded by noise. Thus, some processing of images is required to filter out the noise without losing vital image features. Even though the current literature offers a variety of denoising methods, the fidelity and efficacy of their denoising are sometimes uncertain. Thus, here we propose a novel and computationally efficient image denoising method that is capable of producing accurate images. To preserve image smoothness, this method inputs patches partitioned from the image rather than pixels. Then, it performs denoising on the manifold underlying the patch-space rather than that in the image domain to better preserve the features across the whole image. We validate the performance of this method against benchmark image processing methods.}




\maketitle
\section{Introduction}
\label{sec:introduction}
Modern systems such as satellites and medical imaging instruments rely on cameras operating in diverse environments to capture high-quality images of interest \cite{Joyce2009, Lehmann1999}. Those images are often contaminated with significant noise during acquisition, compression, and transmission that leads to distortion and loss of image information \cite{Fan2019}. Noisy images degrade the performance of subsequent image processing tasks such as image analysis, tracking, and even video processing. Thus, before processing images, an extra step should be performed to denoise images. Since noise, edges, and texture are high-frequency components of an image, distinguishing each component, especially noise, is a non-trivial task \cite{Fan2019}. This limitation frequently causes loss of some vital features of the recovered image. Thus, the recovery of high-quality images without losing vital features is an essential attribute of the denoising process.

Deep learning based image denoising methods, such as \cite{Devalla2018}, \cite{Zhang2017}, and \cite{Zhang2018}, learn a mapping function on a training set that contains clean image pairs by optimizing a loss function \cite{fan2019brief}. Recently, these methods have received surged attention as they have performed well in many computer vision tasks \cite{fan2019brief}. However, the deep learning image denoising frameworks suffer from major drawbacks, that are prominent in typical neural networks, such as difficulties in training when the noise contamination is high, vanishing gradient when the network is considerably deep, and high computational cost due to repeated training \cite{gajamannage2022reconstruction}.

Other than a few deep learning based image denoising frameworks that became existed in recent years, the image denoising literature is significantly dominated by non deep learning based image denoising frameworks. Non deep learning based image denoising methods can be categorized into two types: patch-based and pixels-based. Patch-based image denoising methods, such as \cite{Dabov2007,Elad2006,Lebrun2013,Buades2005,Yan2013,Szlam2008, Zontak2011, Bougleux2009,Taylor2012}, partition an input image into blocks, called patches, and process these patches locally in ``patch-space”  to estimate the true pixel values of the original image \cite{Chatterjee2012}. Patch-based image denoising methods are well known for better performance in contrast to that of pixel-based methods \cite{Alkinani2017}. Patch-based image denoising approaches possess noteworthy advantages such as their efficiently smooth flat regions due to overlaps between patches, and their ability to preserve fine image details and sharp edges \cite{Alkinani2017}. Sparse 3-D transform-domain collaborative filtering \cite{Dabov2007}, abbreviated as BM3D, is the most popular denoising method that consists of two-stage non-local collaborative filtering in the transform domain. BM3D stacks similar patches into 3D groups by block matching and then these 3D groups are transformed into the wavelet domain. Then, hard thresholding or Wiener filtering with coefficients is employed in the wavelet domain followed by applying an inverse transformation of coefficients. Finally, the denoised version of the image is constructed by aggregating all the estimated patches. However, when the noise increases gradually, the denoising performance of BM3D decreases greatly and artifacts are introduced, especially in flat areas \cite{Fan2019}.

Sparse and redundant representations over learned dictionaries \cite{Elad2006}, abbreviated as KSVD, is another popular denoising method that is based on sparse and redundant representations over trained dictionaries. Using the KSVD algorithm, a dictionary that describes the image content is effectively obtained. Since this denoising method is primarily based on the KSVD algorithm, it is named as KSVD denoising (for convenience, we simply call KSVD for this denoising method later). The training is performed either using the corrupted image itself or training on a corpus of high-quality image database. Here, the user needs to threshold the required depth of the sparsity that ensures the sparse representation uses no more than this many columns of the dictionary to reconstruct every image patch instance. Since the KSVD is limited in handling small image patches, authors extend its deployment to arbitrary image sizes by defining a global image prior in a Bayesian reconstruction framework that forces sparsity over patches in every location in the image. However, this method has computational deficiencies since both the choice of an appropriate dictionary for a dataset is a non-convex problem and the implementation of KSVD is an iterative approach that does not guarantee to find the global optimum \cite{Rubinstein2010}. Wavelets denoising with empirical Bayes thresholding, abbreviated as BWD (Bayes wavelet denoising),  \cite{Johnstone2004}, has also shown good denoising performance. Since wavelets localize features of an image to different scales, important signal or image features can be preserved  while removing the noise. Specifically, the wavelet transformation leads to a sparse representation for many real-world images where this representation concentrates image features in a few large magnitude wavelet coefficients. Small wavelet coefficients typically represent noise that can be shrunk or removed without affecting the image quality. Empirical Bayes approach is used as the threshold rule that is based on the assumption that the image has an independent prior distribution given by a mixture model. The inverse wavelet transform is applied on shrunk coefficients to generate the noise-free version of the image. However, such wavelet denoising approaches suffer from high computational cost, less natural denoising, and less robustness.

Here, we propose a novel denoising method that uses eigenvectors of the Gramian matrix of geodesic distances. In our method, first, we partition the noisy image into partially overlapping moving square-patches with a known length, say $\rho$, such that each patch is centered at one unique pixel of the image. Each patch is a point in a $\rho^2$-dimensional space where a low-dimensional manifold underlies \cite{gajamannage2015a, gajamannage2015identifying}. This manifold representation is similar to the wavelet domain representation of BM3D and BWD, and redundant dictionary representation of KSVD, where the image features are concentrated. Revealing this hidden manifold helps better explain the geometry of the patch-set and then helps identify the features in the image. Learning such a manifold is the salient step in the discipline of \emph{Dimensionality Reduction} from where we borrow the basic concept for the proposed method for denoising images. In the context of image processing, the process of projecting the high-dimensional data of the patch-space into a low-dimensional manifold is technically the same as eliminating the noise in the image since extra dimensions mostly represent noise and minor information. This projection is conducted using the Gramian matrix of geodesic distances \cite{SGE,gajamannage2021}. Due to this manifold approach, our method can be considered to be a non-local denoising method that performs denoising in the patch-space rather than that in the image domain to preserve the features across the image.

Geodesics on this manifold are approximated using a custom-made graph structure over the patch-space. This graph structure is made in two steps: 1) we represent all the patches as vertices, 2) for a given neighborhood parameter $\delta$, we search $\delta$ many nearest neighbor patches for each patch and join each pair of neighbors with an edge having the weight equal to the Euclidean distance between them. This neighborhood search constructs a graph structure on the dataset where each point is treated as a vertex. This graph structure closely mimics the geometry of the manifold hidden in the patch-space. Then, we compute the shortest path distance between each pair of vertices as an approximation to its geodesic and construct the geodesic distance matrix. Since nearby points on this manifold represent similar patches of the image regardless of their physical location in the image, geodesic distances encode similarity between patches. Thus, this graph structure approximating geodesic distances is similar to the 3D stacking of the similar patches in BM3D. The advantage of using geodesic distance over the trivial Euclidean distance as a proximity for the manifold distance is that the geodesic distance is nonlinear whereas Euclidean distance is linear. Nonlinear proximity of a manifold mimics the manifold closely so that it helps capturing the true geometry of the manifold underlying the image patch-space. Thus, the adoption of the geodesic distance over the adoption of Euclidian distance ensures better quality of the denoised images that retains essential image features.

The geodesic distance matrix is transformed into its Gramian matrix by double centering. Eigenvalues and eigenvectors of the Gramian matrix of a dataset explain the geometric structures of the dataset \cite{gajamannage2015a, Bahadur2019}. Thus, those eigenvalues and eigenvectors are used to reduce the dimensionality of high-dimensional datasets in Dimensionality Reduction methods such as \cite{gajamannage2021,SGE}. Bigger the eigenvalue of the Gramian, more prominent the features presented by that eigenvalue and its eigenvector. As the noise in an image is less prominent than that of the texture or cartoon, the big eigenvalues and their eigenvectors are dominated by the texture, while the smaller ones mostly represent the noise. Thus, we utilize a subset of fewer eigenvectors of the Gramian matrix to construct the noise-free patches. Usage of fewer prominent eigenvectors of the Gramian matrix is similar to hard thresholding or Wiener filtering on the wavelet domain in BM3D, thresholding to define the sparsity on the dictionary in KSVD, and empirical Bayes thresholding on the wavelet coefficients in BWD. Finally, we merge these noise-free patches to generate the noise-free version of the original noisy image. Since our method performs the denoising task using the Gramian matrix of the shortest graph distances mimicking the manifold geodesics, we name our method as \emph{Geodesic Gramian Denoising} and abbreviate it as GGD.

Our paper is organized as follows: First, we provide the detailed information and theory associated with the development of GGD in Sec.~\ref{sec:method}. Then, we analyze the sensitivity of GGD with respect to its parameters and compare the performance of GGD against benchmark denoising methods with respect to both the input parameters and different levels of corruption in Sec.~\ref{sec:analysis}. Finally, we provide a summary along with conclusions in Sec.~\ref{sec:conclusion}. Table~\ref{tab:nome} and Table~\ref{tab:abrv} state the notations and abbreviations, respectively, used in this article along with their descriptions.

\subsection{Contributions}
Our proposed denoising method, GGD, makes the following contributions to the literature:
\begin{itemize}
\item GGD is a novel denoising algorithm that leverages eigenvectors of the Gramian matrix of geodesic distances  between patches. This geodesic's Gramian approach helps denoising the original noisy image with a small subset of eigenvectors.  
\item GGD is a non-local denoising scheme that performs denoising in the patch-space rather than that in the image domain. Patch-based methods are well known for preserving image smoothness, fine image details, and sharp edges, across the entire image than those of pixel-based methods.
\item In contrast to the general non-local denoising methods that are well known to have many parameters, GGD possesses only three parameters that can easily be pre-determined.
\end{itemize}

\begin{table}[htp]
\caption {Nomenclature} \label{tab:nome}
\begin{tabular}{p{1.2cm}|p{6.5cm}}
Notation &  Description \\

\hline
$d(k,k')$ & Distance between patches $\bs{u}(\bs{x}_{k})$ and $\bs{u}(\bs{x}_{k'})$\\
$L$ & Eigenvector threshold such that $l=1,\dots,L$\\
$k$ & Index of the $ij$-th pixel such that $k=n(i-1)+j$ \\
$\lambda_l$ & $l$-th eigenvalue \\
$n$ & Length and width of the image \\
$\delta$ & Nearest neighbor parameter \\
$\rho$ &  Patch size \\
$\Delta$ & Reconstruction error \\

\hline
$\tilde{\mathcal{U}}$ & Denoised Image\\
$\mathcal{D}$ & Geodesic distance matrix\\
$\mathcal{G}$ & Gramian matrix\\
$\mathcal{I}$ & Original image \\
$I$ & Identity matrix\\
$\mathcal{U}$ & Input image for the algorithm (often noisy) \\
$\Gamma$ & Weights of Shepard's method \\

\hline
$\tilde{\bs{u}}(\bs{x}_{ij})$ & Denoised version of the patch $\bs{u}(\bs{x}_{ij})$\\
$V$ & Eigenvectors of the matrix $\mathcal{G}$ such that $V=[\bs{\nu}_1\vert \dots \vert \bs{\nu}_l \vert \dots \vert \bs{\nu}_{n^2}]^T$\\
$\Lambda$ & Eigenvalues of the matrix $\mathcal{G}$ such that $\Lambda = diag(\lambda_1, \dots, \lambda_l, \dots, \lambda_{n^2})$\\
$\bs{x}_{ij}$ & $ij$-th pixel of the image \\
$\bs{\nu}_l$ & $l$-th eigenvector \\
$\bs{u}(\bs{x}_{ij})$ & Patch centered at the point $\bs{x}_{ij}$\\

\hline
$G(V,E)$ & Graph $G$ with the vertex set $V$ and edge set $E$ \\
$\mathcal{N}(\bs{x}_k)$ &Neighborhood at the pixel $k$ \\

\end{tabular}
\end{table}

\begin{table}[htp]
\caption {Abbreviations} \label{tab:abrv}
\begin{tabular}{p{1.2cm}|p{6.5cm}}
Notation &  Description \\

\hline
BM3D & Sparse 3-D transform-domain collaborative filter denoising \cite{Dabov2007}\\
KSVD & Denoising with sparse and redundant representations over learned dictionaries \cite{Elad2006}\\
BWD & Wavelet denoising with empirical Bayes thresholding \cite{Johnstone2004} \\
GGD & Geodesic Gramian denoising \\
NLB & Nonlocal Bayesian image denoising \cite{Lebrun2013}\\ 
AD & Anisotropic diffusion \cite{Weickert1998}\\
ID & Isotropic diffusion \cite{Perona1990, Bernardes2010} \\
RMSE & Root mean square error \\
PSNR & Peak signal to noise ratio \\
SSIM & Structural similarity index measure \\
\end{tabular}
\end{table}

\section{Geodesic Gramian denoising}\label{sec:method}
In this section, first, we describe the partition of an image into patches as well as the low-dimensional structure of the patch-space. Second, we state both the construction of a graph structure in the patch-space and the approximation of geodesics. Then, we provide the formulation of the Gramian matrix of the geodesic distances. We explain the technique of denoising patches next. Finally, we explain the construction of the noise-reduced version of the input image from the denoised patches.

\subsection{Patch-set}\label{sec:patch-set}
While there are several approaches to partition a noisy image into patches, we partition the given image, denoted as $\mathcal{U}$, of size $n \times n$ into equal-sized square-shaped patches, denoted as $\bs{u}(\bs{x}_{ij})$'s; $i,j=1,\dots,n$, of odd length, denoted as $\rho$, as defined in Definition~\ref{def:patch}. We create square-shaped and odd length patches for the convenience of the formulation of GGD. As we make a patch centered at each pixel, the neighboring patches overlap each other. We replicate the boundary for the required number of times to partition a patch of $\rho \times \rho$ pixels that is centered on a pixel either at the boundary of the image or as close as $\lceil\rho/2\rceil$ pixels to the boundary. This process creates $n^2$ patches for an image of size $n\times n$ that we call the patch-set. Each patch of $\rho \times \rho$ dimensions is treated as a point in a space of $\rho^2$-dimensions; thus, the extrinsic dimensionality of this  patch-set is $\rho^2$. 

\vspace{2mm}
\begin{definition}\label{def:patch}
Let $\bs{x}_{ij}$ be a pixel of the image with horizontal and vertical displacements $i$ and $j$ units, respectively, from the top-left corner of the image domain. We define a square-shaped patch of an odd length  as the square-shaped block of pixels $\bs{u}(\bs{x}_{ij})$ of length $\rho$ centered at $\bs{x}_{ij}$ given by
\begin{equation}\label{eqn:patch}
\bs{u}(\bs{x}_{ij})=\begin{bmatrix}
	u_1(\bs{x}_{ij})\\
	u_2(\bs{x}_{ij})\\
	\vdots\\
	u_{\rho^2}(\bs{x}_{ij})\\
\end{bmatrix}=
\begin{bmatrix}
	\mathcal{U}(i-\lfloor \rho/2\rfloor,j-\lfloor\rho/2 \rfloor)\\
	\mathcal{U}(i-\lfloor \rho/2\rfloor+1,j-\lfloor \rho/2\rfloor)\\
	\vdots\\	
	\mathcal{U}(i+\lfloor \rho/2\rfloor,j+\lfloor \rho/2\rfloor)\\
\end{bmatrix}.
\end{equation}
For simplicity, sometimes we write $\bs{u}(\bs{x}_k)$ for $\bs{u}(\bs{x}_{ij})$ where $k=n(i-1)+j$ and $1\le k \le n^2$.
\end{definition}   
\vspace{2mm}

The space that the path-set represents  is known as the high-dimensional input space in the field of \emph{Dimensionality Reduction} where the geometry of the patch-set is governed by an underlying low-dimensional manifold structure. Dimensionality reduction methods project this high-dimensional data onto this low-dimensional manifold in order to learn the prominent features in the dataset \cite{gajamannage2015a,gajamannage2016}. The extra dimensions that the high-dimensional data possesses than that of the manifold are mostly the non-prominent features including  noise. We use this concept of dimensionality reduction to eliminate noise in the patch-set. For that, first, we define geodesic distance matrix in Sec.~\ref{sec:graph}

\subsection{Geodesic distance matrix}\label{sec:graph}
Geodesics are defined as the true distances on the manifold. However, the computation of such true manifold distances is infeasible due to finite sampling. Thus, we approximate the geodesic distances as the shortest paths on a graph structure that we create in the patch-set. Specifically, we run the neighborhood search algorithm given in \cite{agarwal1999geometric} with a user input neighborhood parameter, defined as $\delta$, on the patch-set. This algorithm searches $\delta$ nearest neighbors for each patch, representing a $\rho^2$-dimensional point, using Euclidean distance. Then, we create a graph structure $G(V,E)$ on this dataset by defining the points, $\{\bs{u}(\bs{x}_k)\vert k=1,\dots,n^2\}$, as vertices, $V$. We define the edge set, $E$, by joining each pair of nearest neighbor points, say $\bs{u}(\bs{x}_{k})$ and $\bs{u}(\bs{x}_{k'})$, with an edge having the weight equal to the Euclidean distance, denoted as $d(k,k')$,
\begin{equation}\label{eqn:eudist}
d(k,k')=\|\bs{u}(\bs{x}_{k})-\bs{u}(\bs{x}_{k'})\|_2,
\end{equation} 
between them. 

The geodesic distance between two points in the dataset or two patches in the patch-set is approximated as the \emph{shortest path distance} between the corresponding two vertices in the graph $G(V,E)$. The shortest path distance between any two vertices can be computed in many ways, including Dijkstra's algorithm \cite{dijkstra1959note}. However, we employ Floyd's algorithm \cite{floyd1962algorithm} for this task as it computes the shortest paths between all the pairs of vertices in one batch, and is more efficient than Dijkstra's algorithm in this case. We approximate all the geodesic distances between patches and formulate the geodesic distance matrix $\mathcal{D} \in\mathbb{R}^{n^2\times n^2}_{\ge0}$ of the patch-set. Then, we transform this geodesic distance matrix into its Gramian matrix in Sec.~\ref{sec:gramian}.

\subsection{Gramian matrix}\label{sec:gramian}
We transform the geodesic distance matrix $\mathcal{D} \in\mathbb{R}^{n^2\times n^2}_{\ge0}$ into its Gramian matrix, denoted by $\mathcal{G}_{n^2 \times n^2}$, using
\begin{equation}\label{eqn:gram}
\mathcal{G}[i,j]=-\frac{1}{2}\big[\mathcal{D}[i,j]-\mu_i(\mathcal{D}) -\mu_j(\mathcal{D})+\mu_{ij}(\mathcal{D})\big],
\end{equation} 
where $\mu_i(\mathcal{D})$, $\mu_j(\mathcal{D})$, and $\mu_{ij}(\mathcal{D})$ are the means of the $i$-th row of the matrix $\mathcal{D}$, $j$-th column of that matrix, and the mean of the full matrix, respectively, \cite{lee2004nonlinear}. Gramian matrix of geodesic distances is \emph{real-valued}, \emph{symmetric}, and \emph{positive semi-definite}, see Definition~\ref{def:psd} \cite{lee2004nonlinear}.

\begin{figure*}[htp]
	\begin{center}
	\includegraphics[width=1\textwidth]{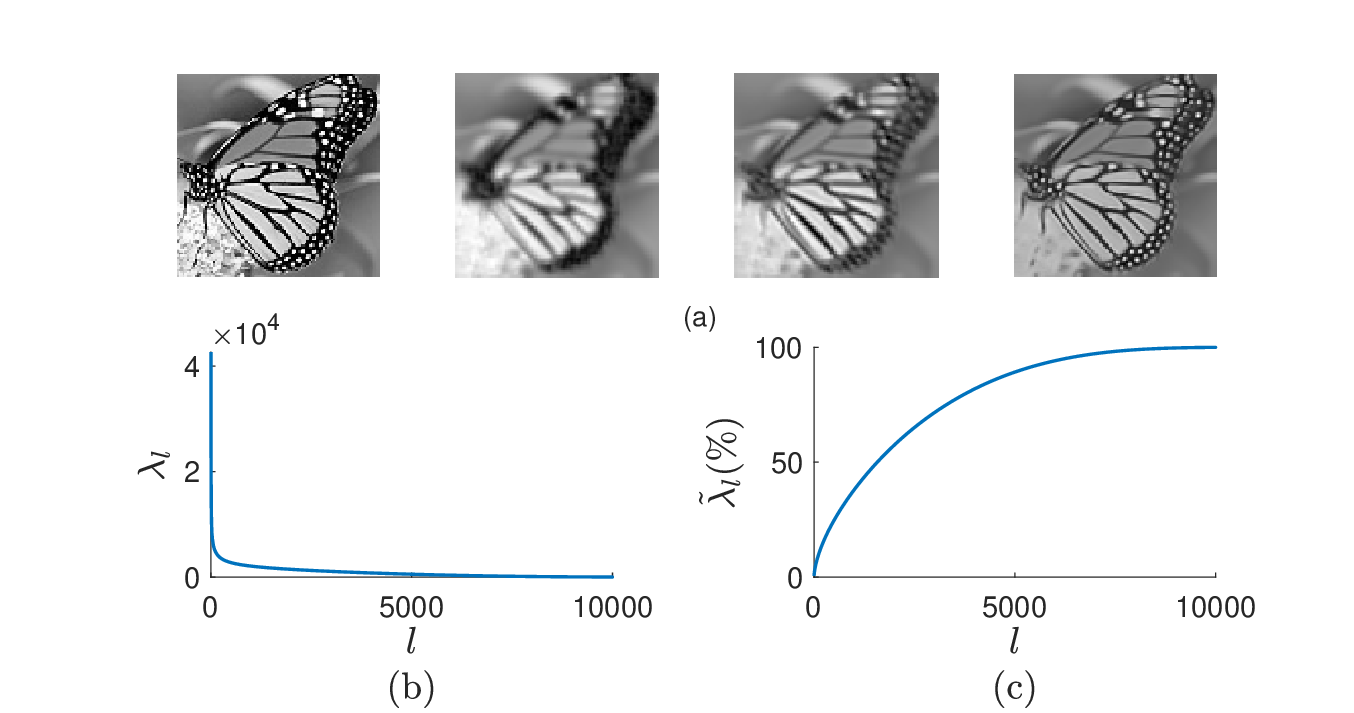}
	\end{center}
	\caption{Information retention of eigenvalues and eigenvectors of the Gramian matrix constructed for an image of size 100$\times$100. First, we construct the patch-set of $100^2$ patches, each with the length $\rho=5$, from the leftmost image in (a). Second, we create a graph structure in the patch-space with the neighborhood size $\delta=10$ and approximate the geodesic distances using Floyd's algorithm. Third, we formulate the Gramian matrix of the geodesic distances. We reconstruct the leftmost image in (a) using the eigenvector associated with the biggest eigenvalue, see the second image from the left of (a), using the eigenvectors associated with the five biggest eigenvalues, see the third image from the left of (a), and using the eigenvectors associate with the 100 biggest eigenvalues, see the rightmost image in (a). (b) The magnitude of eigenvalues, denoted by $\lambda_l$ for $l=1,\dots,n^2$, from the biggest to the smallest, versus eigenvalue index, denoted by $l$, of the Gramian matrix and, (c) cumulative percentage of eigenvalues, denoted by $\tilde{\lambda}_l$ for $l=1,\dots,n^2$, versus the index.}
	\label{fig:gram}
\end{figure*}

The eigenvalues and eigenvectors of the Gramian matrix can be used to describe the properties of the underlying manifold of the patch-set and then that manifold can be used to describe the features of the image. While big eigenvalues and their corresponding eigenvectors of the Gramian matrix represent prominent features of the image including edges and corners of the texture, small eigenvalues and eigenvectors of that represent non-prominent features such as fine image details and noise. This statement is supported by the observations in Fig~\ref{fig:gram} where we analyze the information retention of the eigenvalues and eigenvectors of the Gramian matrix. Therein, we produce the Gramian matrix of an image of size $100\times 100$ as described in Secs.~\ref{sec:patch-set}, \ref{sec:gramian}, and \ref{sec:graph}. Fig.~\ref{fig:gram}(a) shows that only a few eigenvectors, out of $100^2$, of the Gramian matrix is sufficient to reconstruct the image to a good level of quality. Figs.~\ref{fig:gram}(b) and \ref{fig:gram}(c) show that only a few eigenvalues of the Gramian matrix retain most of the information of the image, while most of them retain less information. Consider that the denoising of patches that we will present in Sec.~\ref{sec:denoi_patches} was not performed for this example; however, the technical details of the image reconstruction using eigenvectors will be provided in Sec.~\ref{sec:merging_patches}.

\vspace{2mm}
\begin{definition}\label{def:psd}
Let $\mathcal{G}$ be a square-shaped matrix of order $n^2\times n^2$. The eigenvalue decomposition of $\mathcal{G}$ is
\begin{equation}
\mathcal{G}=V\Lambda V^T,
\end{equation}
where $V=[\bs{\nu}_1\vert \dots \vert \bs{\nu}_l \vert \dots \vert \bs{\nu}_{n^2}]^T$ is a matrix that represents eigenvectors $\bs{\nu}_l$'s by its rows and $\Lambda = diag(\lambda_1, \dots, \lambda_l, \dots, \lambda_{n^2})$ is a diagonal matrix that represents eigenvalues $\lambda_l$'s. The matrix $\mathcal{G}$ is positive semi-definite, if and only if $\lambda_l\ge0$ for all $l$.
\end{definition}
\vspace{2mm}

\subsection{Denoising patches}\label{sec:denoi_patches}
The patches are denoised using only a few, say $L$ (eigenvector threshold), prominent eigenvectors of the Gramian matrix as they represent essential features of the image. For $l=1,\dots,n^2$ and $\lambda_1\ge \dots \ge \lambda_l \ge \dots \lambda_{n^2}$, $(\lambda_l, \bs{\nu}_l)$ represents eigenvalue and eigenvector pairs of the Gramian matrix. We denote the noise-reduced version of the patch $\bs{u}_k$ as $\tilde{\bs{u}}_k$ that we produce by
\begin{equation}\label{eqn:denoise}
\tilde{\bs{u}}(\bs{x}_k) = \sum^{L}_{l=1} \langle \bs{u}(\bs{x}_k), \bs{\nu}_l \rangle \bs{\nu}_l,
\end{equation}
where $l=1,\dots,L$. Here, $\langle \cdot, \cdot \rangle$ denotes the inner product according to Definition~\ref{def:inner_product}. Note that this $k$ is related to row index $i$ and the column index $j$, both measured from the top-left corner of the image, by $k=n(i-1)+j$. Denoised patches are merged using Shepard’s method as stated in Sec.~\ref{sec:merging_patches}.

\vspace{2mm}
\begin{definition}\label{def:inner_product}
Let $\bs{\nu}^{(1)}=\left(\nu^{(1)}_1,\dots, \nu^{(1)}_l , \dots, \nu^{(1)}_{n^2}\right)$ and $\bs{\nu}^{(2)}=\left(\nu^{(2)}_1, \dots, \nu^{(2)}_l, \dots, \nu^{(2)}_{n^2}\right)$ be two vectors, the inner product of these vectors is defined as
\begin{equation}
\langle \bs{\nu}^{(1)}, \bs{\nu}^{(2)} \rangle = \sum^{n^2}_{l=1} \nu^{(1)}_l \nu^{(2)}_l.
\end{equation}
\end{definition}
\vspace{2mm}

\subsection{Merging denoised patches}\label{sec:merging_patches}
In our approach, each pixel in the image domain is overlapped with $\rho^2$ patches. This overlapping makes each pixel location in the image also exist in nearby $\rho^2$ patches. These nearby pixels are within the radius of $\rho/2$ units from the target pixel, say $\bs{x}_k$. We denote this neighborhood as $\mathcal{N}(\bs{x}_k)$ and define as
\begin{equation}
\mathcal{N}(\bs{x}_k)=\{\bs{x}_{t} \ \vert \ \ \|\bs{x}_k-\bs{x}_{t}\|_{\infty}\le \rho/2\},
\end{equation}
\cite{Meyer2014}. 

In order to reconstruct the image intensity at a pixel location of the noise-reduced version of the image, we have to get an estimate from the same pixel location in $\rho^2$ nearby patches. For each pixel $\bs{x}_t\in\mathcal{N}(\bs{x}_k)$, there exists a new index $t_n$ such that the extrinsic pixel location ($i,j$) at that new index of the patch $\tilde{\bs{u}}(\bs{x}_t)$, denoted by $[\tilde{\bs{u}}(\bs{x}_t)]_{t_n}$, is the same as the extrinsic pixel location of $\bs{x}_k$. We use this new index in the patch merging step below.  We combine all these estimates using a moving least square approximation given by Shepard’s method \cite{Shepard1968} and construct the pixel $\bs{x}_k$ of the denoised version of the image as
\begin{equation}\label{eqn:merge}
\tilde{\mathcal{U}}(\bs{x}_k)=\sum_{\bs{x}_t \in \mathcal{N}(\bs{x}_k)} \Gamma (\bs{x}_k,\bs{x}_t) [\tilde{\bs{u}}(\bs{x}_t)]_{t_n},
\end{equation}
where the weights $\Gamma(\bs{x}_k,\bs{x}_t)$ are defined as 
\begin{equation}\label{eqn:shepard_weight}
\Gamma(\bs{x}_k,\bs{x}_t)=\frac{e^{-\|\bs{x}_k-\bs{x}_t\|^2}}{\sum_{\bs{x}_{t'}\in \mathcal{N}(\bs{x}_k)} e^{-\|\bs{x}_k-\bs{x}_{t'}\|^2}}.
\end{equation}
The weighting term in Eqn.~\eqref{eqn:shepard_weight} weights close by pixels with more weight while the faraway pixels with less weight. Thus, according to Eqn.~\eqref{eqn:merge}, merging assures that the pixel $\bs{x}_k$ of the reconstructed image is highly influenced by the pixels at the same location of the nearby patches. The main steps of GGD are summarized in Algorithm~\ref{alg:algorithm}.

\begin{algorithm}[!ht]
\caption{ \textit{Geodesic Gramian Denoising (GGD).\\
\begin{tabular}{l p{10cm}}
Inputs: & noisy image ($\mathcal{U}_{n\times n}$), patch length ($\rho$), nearest neighborhood size ($\delta$), and eigenvector threshold ($L$).\\
Outputs: & noise-reduced image ($\tilde{\mathcal{U}}_{n\times n}$).\\
\end{tabular}
}}
\begin{algorithmic}[1]
\State  Construct $n^2$ overlapping square-shaped patches each with the length $\rho$ from the noisy image $\mathcal{U}_{n\times n}$ and denote the patch-set as $\{\bs{u}(\bs{x}_k)\vert k=1,\dots,n^2\}$, (Sec.~\ref{sec:patch-set}).
	
\State Produce the graph structure $G(V,E)$ from the patch-set using the nearest neighbor search algorithm in \cite{agarwal1999geometric}. Use Floyd's algorithm in \cite{floyd1962algorithm}, to approximate the geodesic distances in the patch-space and then produce the geodesic distance matrix $\mathcal{D}$, (Sec.~\ref{sec:graph}).
	
\State Construct the Gramian matrix $\mathcal{G}$ from the geodesic distance matrix $\mathcal{D}$ using Eqn.~\eqref{eqn:gram}, (Sec.~\ref{sec:gramian}).
	
\State Compute the eigenvectors $\{\nu_l\vert l = 1,\dots L\}$ corresponding to the $L$ biggest eigenvalues of the Gramian matrix $\mathcal{G}$ and use Eqn.~\eqref{eqn:denoise} to produce noise-free patches $\{\tilde{\bs{u}}(\bs{x}_k)\vert k=1,\dots,n^2\}$, (Sec.~\ref{sec:denoi_patches}).
	
\State Merge noise-free patches using Eqns.~\eqref{eqn:merge} and \eqref{eqn:shepard_weight}, and generate the denoise image $\tilde{\mathcal{U}}_{n\times n}$, (Sec.~\ref{sec:merging_patches})
\end{algorithmic}\label{alg:algorithm}
\end{algorithm}

\section{Performance analysis}\label{sec:analysis}
We analyze the performance of GGD by both visual perception and three similarity metrics, namely, Root Mean Square Error (RMSE), Peak Signal to Noise Ratio (PSNR), and Structural Similarity Index Measure (SSIM). First, we  analyze the sensitivity of GGD to both the noise contamination ($\epsilon$) of an input image and the parameters of GGD, namely, patch size ($\rho$), neighborhood size ($\delta$), and eigenvector threshold ($L$). Then, we compare the performance of GGD with six benchmark image denoising methods.

RMSE, see Def.~\ref{def:rmse}, ranges  between $0$ and $\infty$, is a primary measure to detect the numerical trad-off of an image from a reference image, e.g, while the original noise-free image is the reference image, the other image is a denoised approximation of a noisy version of the original image \cite{hore2010}. PSNR, see Def.~\ref{def:psnr}, measures the numerical difference of an images from a reference image, with respect to the maximum possible pixel value of the reference image \cite{hore2010} where PSNR ranges between $0$ and $\infty$. If the original noise-free image is the reference image and a denoised approximation of its noisy version is the other image, a higher PSNR value provides better quality of the approximation since it approaches infinity as the RMSE approaches zero \cite{hore2010}. SSIM, see Def.~\ref{def:ssim}, ranging between -1 and 1 measures the structural similarity of an image to a reference image \cite{Wang2004}. Ideal approximation provides 1 for SSIM since such approximation ensures $\mu_{\mathcal{I}}=\mu_{\tilde{\mathcal{U}}}$, $\sigma_{\mathcal{I}}=\sigma_{\tilde{\mathcal{U}}}$, and $\sigma_{\mathcal{I}\tilde{\mathcal{U}}}=\sigma_{\mathcal{I}}\sigma_{\tilde{\mathcal{U}}}$ in Def.~\ref{def:ssim}. Instead of using traditional error summation methods, SSIM is designed by modeling any image distortion as a combination of three factors, namely, luminance distortion, contrast distortion, and loss of correlation. 

\vspace{2mm}
\begin{definition}\label{def:rmse}
Let, two-dimensional matrix $\mathcal{I}$ represents a reference image of size $n\times n$ and $\tilde{\mathcal{U}}$ represents any other image of interest. Root Mean Square Error \cite{hore2010}, abbreviated as RMSE, of the image $\tilde{\mathcal{U}}$ with respect to the reference image $\mathcal{I}$ is defined as  
\begin{equation}
RMSE(\mathcal{I},\tilde{\mathcal{U}}) = \sqrt{\frac{\sum_{(i,j)\in\mathbb{N}^{n\times n}} (\mathcal{I}[i,j]-\tilde{\mathcal{U}}[i,j])^2}{n^2}}.
\end{equation}
\end{definition}   
\vspace{2mm}

\begin{definition}\label{def:psnr}
Let, two-dimensional matrix $\mathcal{I}$ represents a reference image of size $n\times n$ and $\tilde{\mathcal{U}}$ represents any other image of interest. Peak Signal to Noise Ratio \cite{hore2010}, abbreviated as PSNR, of the image $\tilde{\mathcal{U}}$ with respect to the reference image $\mathcal{I}$ is defined as  
\begin{equation}
PSNR(\mathcal{I},\tilde{\mathcal{U}}) = 20 \log_{10}\left(\frac{\max(\mathcal{I})}{RMSE(\mathcal{I},\tilde{\mathcal{U}})}\right).
\end{equation}
Here, $\max(\mathcal{I})$ represents the maximum possible pixel value of the image $\mathcal{I}$. Since the pixels in our images of interest are represented in 8-bit digits, $\max(\mathcal{I})$ is 255.
\end{definition}   
\vspace{2mm}

\begin{definition}\label{def:ssim}
Let, two-dimensional matrix $\mathcal{I}$ represents a reference image of size $n\times n$ and $\tilde{\mathcal{U}}$ represents an image of interest. Structural Similarity Index Measure \cite{Wang2004}, abbreviated as SSIM, of the image $\tilde{\mathcal{U}}$ with respect to the reference image $\mathcal{I}$ is defined as the product of luminance distortion ($I$), contrast distortion ($C$), and loss of correlation ($S$), such as
\begin{equation}
SSIM(\mathcal{I},\tilde{\mathcal{U}}) = I(\mathcal{I},\tilde{\mathcal{U}}) \ C(\mathcal{I},\tilde{\mathcal{U}}) \ S(\mathcal{I},\tilde{\mathcal{U}}),
\end{equation}
where
\begin{equation}
\begin{split}
I(\mathcal{I},\tilde{\mathcal{U}}) = \frac{2\mu_{\mathcal{I}}\mu_{\tilde{\mathcal{U}}}+c_1}{\mu^2_{\mathcal{I}}+\mu^2_{\tilde{\mathcal{U}}}+c_1},\\
C(\mathcal{I},\tilde{\mathcal{U}}) = \frac{2\sigma_{\mathcal{I}}\sigma_{\tilde{\mathcal{U}}}+c_2}{\sigma^2_{\mathcal{I}}+\sigma^2_{\tilde{\mathcal{U}}}+c_2},\\
S(\mathcal{I},\tilde{\mathcal{U}}) = \frac{\sigma_{\mathcal{I}\tilde{\mathcal{U}}}+c_3}{\sigma_{\mathcal{I}}\sigma_{\tilde{\mathcal{U}}}+c_3}.\\
\end{split}
\end{equation}
Here, $\mu_{\mathcal{I}}$ and $\mu_{\tilde{\mathcal{U}}}$ are means of $\mathcal{I}$ and $\tilde{\mathcal{U}}$, respectively; $\sigma_{\mathcal{I}}$ and $\sigma_{\tilde{\mathcal{U}}}$ are standard deviations of $\mathcal{I}$ and $\tilde{\mathcal{U}}$, respectively; and  $\sigma_{\mathcal{I}\tilde{\mathcal{U}}}$ is the covariance between $\mathcal{I}$ and $\tilde{\mathcal{U}}$. Moreover, $c_1$, $c_2$, and $c_3$ are very small positive constants to avoid the case of division by zero.
\end{definition}   
\vspace{2mm}
\subsection{Sensitivity analysis}\label{sec:sen}
Here, first, we analyze the sensitivity of GGD with respective to the parameters patch size ($\rho$) and neighborhood size ($\delta$). Then, we analyze the sensitivity of GGD with respect to the corruption level ($\epsilon$) of the input images followed by analyzing the sensitivity of GGD with respect to the parameter eigenvector threshold ($L$).

\subsubsection{Patch size and neighborhood size}
We use an image of Lena Forsén, which is commonly used for evaluating image processing algorithms, to analyze the influence of patch size and neighborhood size for the performance. Since the original image is colored, we transform it into a gray image by taking the average across three color channels. The original noise-free image, that we denote by $\mathcal{I}$, is of the size $100 \times 100$. We make three noisy versions, denoted by $\mathcal{U}_{1,2,3}$, of this gray image by imposing additive Gaussian noise with three different levels of standard deviations, $\epsilon=$, 40, 60, and 80, such that
\begin{equation}\label{eqn:noise}
\mathcal{U}=\mathcal{I}+ \ \mathcal{N}(0,\epsilon^2),
\end{equation}
where $\mathcal{N}(0,\epsilon^2)$ denotes a Gaussian random distribution with mean 0 and standard deviation $\epsilon$. Since $\mathcal{U}$'s represent images, the values of the pixels in them should be between 0--255. Thus, we adjust $\mathcal{U}$ by replacing the values less than zero with zeros and the values more than 255 with 255's.
\begin{figure*}[htp]
	\begin{center}
	\includegraphics[width=1\textwidth]{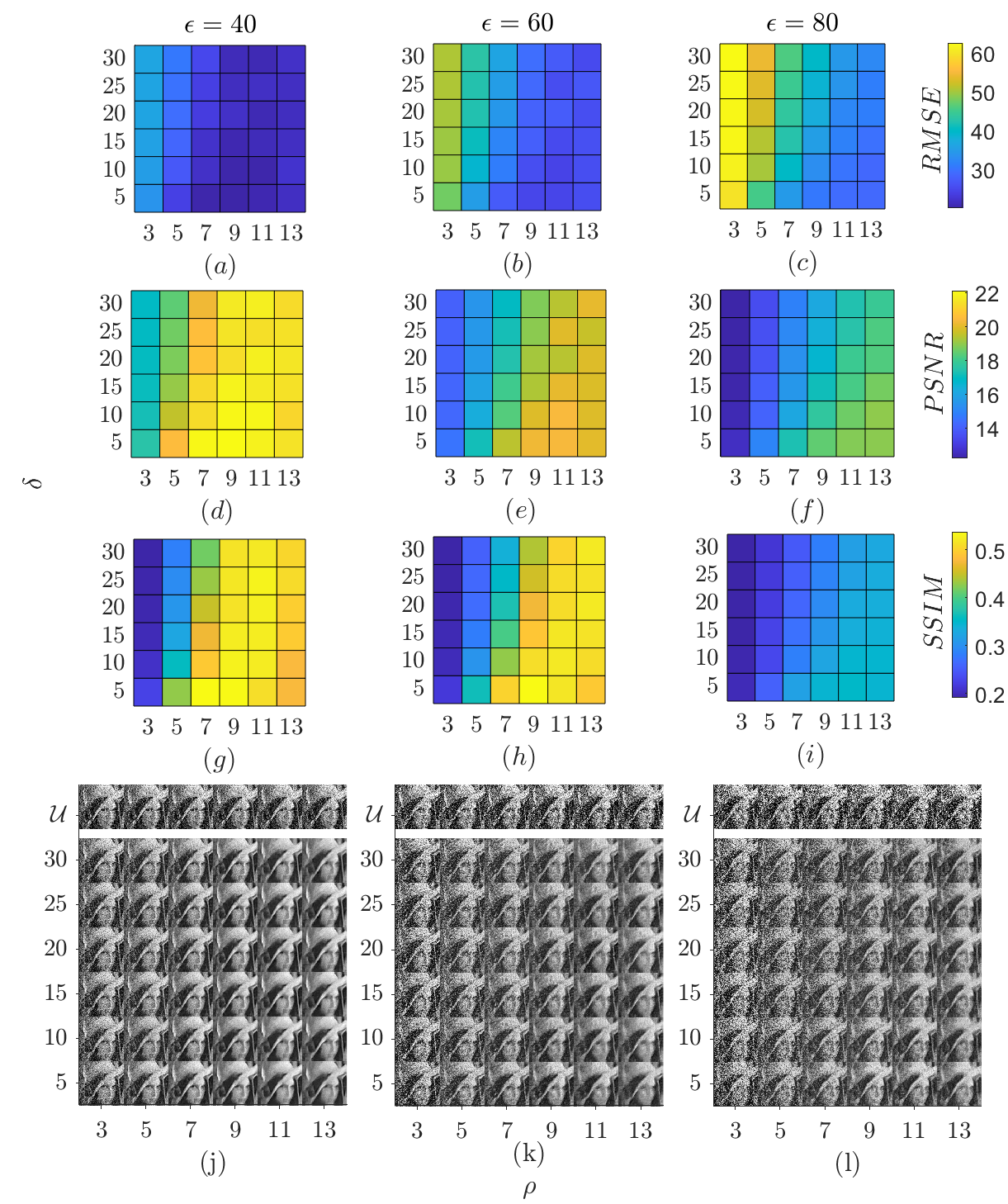}
	\end{center}
	\vspace{-2mm}
	\caption{Denoising performance of GGD with respect to patch size ($\rho$) and neighborhood size ($\delta$). We produce three noisy images from an image of Lena Forsén of size $100\times 100$ by imposing three levels of noise sampled from the Gaussian distributions $\mathcal{N}(0,\epsilon^2)$ where $\epsilon=40$, 60, and 80. We set the eigenvector threshold to an arbitrary 50 and run GGD with 36 pairs of parameter values $\{(\rho,\delta)\vert \ \rho = 3,5,7,9,11,13; \delta = 5,10,15,20,25,30\}$ on each of the three images. Then, we compute the reconstruction errors in terms of RMSE, PSNR, and SSIM of GGD associated with 36 parameter pairs and three noise levels $\epsilon=40, 60, 80$. We run the same experiment four more times each with a different seed (each contributes a sample realization) in the Gaussian distribution that generates additive noise for the corrupted. For each metric, (a-c) RMSE, (d-f) PSNR, and (g-i) SSIM, and for each noise level $\epsilon = 40, 60$, and 80, we average the corresponding performance values over the five sample realizations. Each color-bar is scaled to interpret the values in all the three cases $\epsilon = 40, 60$, and 80 in the same row. We also compute the corruption level of the three noisy images using the same metric where $RMSE=38.76$, $PSNR=16.36$, and $SSIM=.3537$ for $\epsilon=40$; $RMSE=53.87$, $PSNR=13.50$, and $SSIM=.2444$ for $\epsilon=60$; $RMSE=66.82$, $PSNR=11.63$, and $SSIM=.1820$ for $\epsilon=80$. (j-l) The appearance of the noisy images and their denoised images for all the 3$\times$36 cases (see supplementary materials for the enlarged images).}
	\label{fig:rhovsk}	
\end{figure*}

We run our denoising method 36 times with 36 pairs of parameter values ($\rho,\delta$) where $\rho=$ 3, 5, 7, 9, 11, 13 and $\delta= $ 5, 10, 15, 20, 25, 30 on each of the three noised versions of the image. We set the eigenvector threshold arbitrarily as $L=50$ in GGD and compute the noise-reduced version $\tilde{\mathcal{U}}$ of the input image. We run this experiment four more times each with a different random seed (each random seed contributes one sample realization) of the Gaussian distribution in Eqn.~\eqref{eqn:noise} to avoid the random effect. For each pair of parameters and for each sample realization, the reconstruction error is computed using the three similarity metrics, RMSE (Def.~\ref{def:rmse}), PSNR (Def.~\ref{def:psnr}), and SSIM (Def.~\ref{def:ssim})  with the noise-free original image as the reference image  and the denoised image as the image of interest. For each parameter pair, one value for each metric is computed after averaging the five values obtained for different sample realizations. The corruption levels of the input noisy images are also assessed in terms of the same similarity metrics with the noise-free original image as the reference image and the noisy images as the images of interest. For each corruption level and for each metric, we average the metric values obtained across five different sample realizations and compute a single value.

The corruption levels of the input noisy images are $RMSE=38.76$, $PSNR=16.36$, and $SSIM=.3537$ for $\epsilon=40$; $RMSE=53.87$, $PSNR=13.50$, and $SSIM=.2444$ for $\epsilon=60$; $RMSE=66.82$, $PSNR=11.63$, and $SSIM=.1820$ for $\epsilon=80$. Figs.~\ref{fig:rhovsk}(a-i) show that GGD performs better in denoising the image for all the 36 parameter pairs and all the three noise levels than the corresponding input corruption level. Moreover, we observe that the denoising performance can be made substantially better with some parameter values. We observe that the reconstruction error increases when the noise increases. We also observe in Figs.~\ref{fig:rhovsk}(a-i) that when the corruption levels are $\epsilon=40$, 60, and 80, the best denoise performances, in terms of all the three metrics, are observed at $\rho=9, 11$, $\rho=11, 13$, and $\rho\ge 13$, respectively.  Figs.~\ref{fig:rhovsk}(j-l) provides visual evidence for the above trade-offs between performance and parameters where we see that the best denoising is attained around aforesaid $\rho$'s for aforesaid noise levels. This is because if the input image possesses high noise, big patches help smooth the image more, whereas bigger patches might smooth the image too much so that underlying image features might also be distorted. The denoising performance slightly decreases, in terms of all the metrics, when the neighborhood size increases. The reason for that is big neighborhood sizes add more edges into the graph structure $G(V,E)$ presented in Sec.~\ref{sec:graph} which may cause underestimation of the true geodesic distances on the manifold.

\subsubsection{Corruption level of the input image}\label{sec:corr_level}
Here, we analyze the performance of GGD with respect to the corruption level of the input image. We vary the corruption level of images by imposing different levels of Gaussian random noise into the image of Lena Forsén of size $100\times 100$. For that, we vary $\epsilon = 0, 10, 20, \dots, 100$ in Eqn.~\eqref{eqn:noise} and create 10 noisy images. We run GGD over these noisy images with arbitrary parameter values ($\rho, \delta, L$) where $\delta=10$, $\rho=3, 7$, and $L=10$, $10^2$, $10^3$, $10^4$, and denoise them. We compute RMSE (Def.~\ref{def:rmse}), PSNR (Def.~\ref{def:psnr}), and SSIM (Def.~\ref{def:ssim}) for each denoised image with the noise-free original image as the reference image  and the denoised image as the image of interest. Moreover, we compute the corruption level of each noisy image using the same three metrics, with the noise-free original image as the reference image and the noisy images as the images of interest, and use that to compare the denoising results of GGD. Since we added Gaussian random noise onto images, to eliminate the random effect on the results, we run the same experiment 10 times, e.i. realizations, each with a different seed (each seed contributes one sample realization) in the Gaussian random number generator. We average the results over the 10 realizations for each combination of parameters.

\begin{figure*}[htp]
	\begin{center}
	\includegraphics[width=1\textwidth]{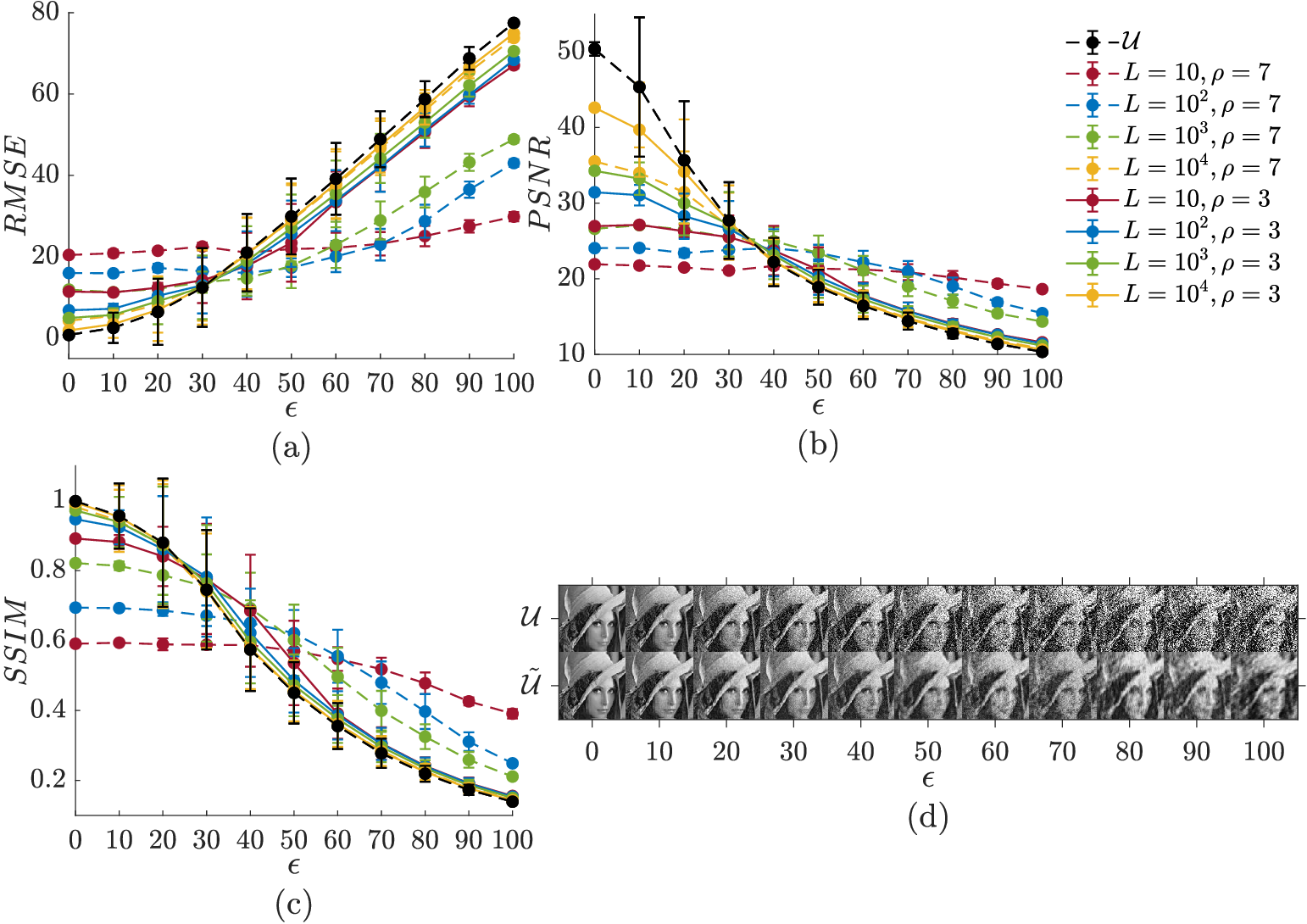}
	\end{center}
	\caption{Image denoising performance of GGD with respect to the intensity of noise contamination ($\epsilon$). We produce 11 images by imposing a variable noise into an image of Lena Forsén of size $100\times 100$ that is sampled from the Gaussian distributions $\mathcal{N}(0,\epsilon^2)$ where $\epsilon=$ 0, 10, $\dots$, 100. We denoise these images using GGD with the parameters ($\rho, \delta, L$) for $\rho=3,7$; $\delta=10$; and $L=10, 10^2, 10^3, 10^4$ where $\delta$, $\rho$, and $L$ denote neighborhood size, patch size, and eigenvector threshold, respectively. We compute the reconstruction errors using the three metrics RMSE, PSNR, and SSIM for all the denoised versions of the noisy images. We run GGD with each of the parameter combinations nine more times with different random seeds (each random seed provides a sample realization) in the Gaussian distribution. For each parameter combination, the means of the reconstruction errors across 10 sample realizations with respect to $\epsilon$ is presented in (a-c) where errorbars represent the standard deviations of the reconstruction errors computed across 10 sample realizations. The corruption levels of the input images, denoted by $\mathcal{U}$, with respect to the noise levels is also computed using the three metrics. We use the corruption levels to compare the denoising results of GGD. (d) The appearance of the noisy images and their denoised images for all the 11 noisy images (see supplementary materials for the enlarged images).}
	\label{fig:noise}
\end{figure*}

Fig.~\ref{fig:noise}(a-c) represent, 1) means of the three reconstruction error metrics RMSE, PSNR, and SSIM of the denoised images across 10 realizations  with respect to the noise levels ($\epsilon$'s) for each combination of the parameters ($\rho,L$) where $\rho=$ 3,7 and $L=10, 10^2, 10^3, 10^4$; and 2) the means of the same three metric values of the noisy images across 10 realizations with respect to the noise levels. The errorbars in the figures represent the standard deviation of the reconstruction errors across 10 realizations. We observe that the errorbars of the noisy images are bigger than that of the denoised images associated with small eigenvector thresholds due to the fact that small eigenvector thresholds remove more noise that will eventually reduce the influence of the random seed to the performance.

Fig.~\ref{fig:noise}(a-c) shows that GGD performs substantially better than the initial noisy image under all the noise levels for at least one of the combination of parameters ($\rho, L$). Moreover, we observe that GGD performs better with bigger patch sizes for higher noise levels whereas it performs better with smaller patch sizes for lower noise levels. This is because bigger patches have more overlapping pixels between nearby patches which makes a high dependency between points on the underlying manifold where it is capable of removing more noise in an image. We also observe that while smaller eigenvector thresholds increase denoising performance under higher noise levels, bigger eigenvector thresholds increase denoising performance under lower noise levels. More eigenvectors retain more features in the denoising of the images with low noise so that more eigenvectors increase denoising performance under low noise levels; however, more eigenvectors also retain more noise if the noise contamination is significant in the noisy images that will subsequently reduce the denoising performance. We also observe that the reconstruction errors of the denoising converge to that of the initial noisy image when all the eigenvectors, $10^4$, are used for the denoising due to the fact that all the eigenvalues essentially produces back the initial noisy image. 

\subsubsection{Eigenvector threshold} 
\begin{figure*}[htp]
	\begin{center}
	\includegraphics[width=1\textwidth]{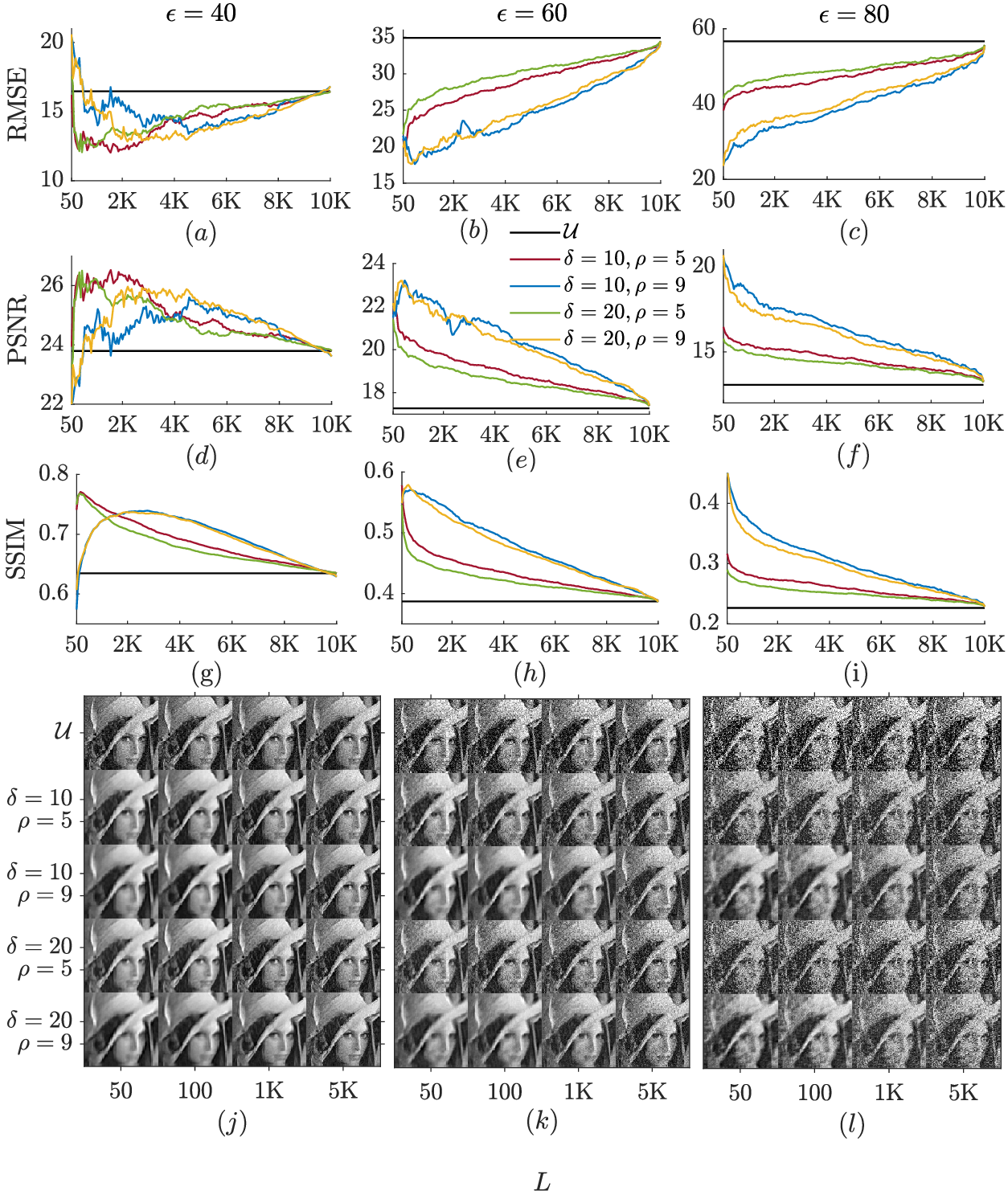}
	\end{center}
	\vspace{-2mm}
	\caption{Denoising performance of GGD with respect to different eigenvector thresholds ($L$'s). We impose random noise sampled from the Gaussian distributions $\mathcal{N}(0,\epsilon^2)$,  where $\epsilon=$ 40, 60, and 80, into the image ``Lena Forsén'' of size $100\times 100$, and generate three noisy test images. Each noisy image is denoised using GGD with four combinations of the parameters neighborhood size ($\delta$) and patch size ($\rho$) such that ($\delta$, $\rho$) = (10,5), (10,9), (20,5), and (20,9) for a sequence of eigenvector thresholds $L=50,100,\dots,10000$ (10000 is the total number of eigenvectors of an image of size $100\times 100$). The reconstruction error of the denoising is computed using three performance metrics RMSE, PSNR, and SSIM that we present in (a-i). Consider that K's in these plots stand for thousand (e.i, $'$000). The corruption levels of the three test images, denoted by $\mathcal{U}$'s, are also computed using the same three performance metrics and used them to compare the denoising performance of GGD. In (a-i), even though the amount of corruption computed using any of the performance metrics for any noisy images is both a scalar and independent of the eigenvector threshold parameter, we represent it as a horizontal line at the metric value across eigenvector thresholds to provide a better comparison. (j-l) The appearance of the noisy images and their denoised images for three noise levels, four parameter combinations, and four eigenvector thresholds (see supplementary materials for the enlarged images).}
	\label{fig:ev}
\end{figure*}

Here, we analyze the influence of the eigenvector threshold on the performance of GGD. We create three images of Lena Forsén of size $100\times 100$ using Eqn.~\eqref{eqn:noise} with three corruption levels $\epsilon$ = 40, 60, and 80. We run GGD over each image with four different parameter sets ($\delta$, $\rho$) = (10,5), (10,9), (20,5), and (20,9) that we chose arbitrarily to generate the denoising at a sequence of eigenvector thresholds such  that  $L=50,100,\dots,10000$. The reconstruction errors of all the denoised images are computed using three reconstruction performance metrics RMSE (Def.~\ref{def:rmse}), PSNR (Def.~\ref{def:psnr}), and SSIM (Def.~\ref{def:ssim}) by treating the noise-free original image as the reference image  and the denoised image as the image of interest. Figs.~\ref{fig:ev} (a-i) show the reconstruction errors computed using three performance metrics RMSE, PSNR, and SSIM for all the 12 experiments. We compute the corruption levels of the input noisy images using the same three performance metrics and compare GGD's denoising performance with them. 

We observe that at some eigenvector thresholds the denoising performance with respect to all the three reconstruction performance metrics is substantially better than the initial corruption for all the three noise levels and all the four parameter combinations. We also observe that the reconstruction error  converges to the corruption level when the eigenvector threshold reaches to its maximum due to the fact that GGD with all the eigenvectors includes not only all the features of the initial noisy image but also all its noise. Independent to the eigenvector threshold, bigger patch sizes improve denoising performance with respect to all the three metrics when the noise contamination is high. However, only for some eigenvector thresholds, smaller patch sizes improve denoising performance with respect to all the three metrics when the noise contamination is low. Neighborhood size doesn't have much influence on denoising performance. The best eigenvector threshold is immensely sensitive to the noise contamination of the image and parameter values in use. We observe that GGD performs significantly better denoising with respect to all the three performance metrics with fewer eigenvector thresholds. Moreover, we observe that the optimum eigenvector thresholds for the best denoising with respect to three performance metrics are close enough. These fewer optimum eigenvectors can be computed using some algebraic techniques rather than working on the entire data matrix that we will discuss with details in Sec.~\ref{sec:conclusion}. 

\subsection{Comparison of GGD with benchmark image denoising methods}\label{sec:other_meth}
After the detailed sensitivity analysis of the parameters in GGD and the analysis of noise contamination of images, as presented in Sec.~\ref{sec:sen}; here, we compare the performance of GGD with six benchmark denoising methods. Those methods are sparse 3-D transform-domain collaborative filtering (BM3D) \cite{Dabov2007}, sparse and redundant representations over learned dictionaries (KSVD) \cite{Elad2006}, wavelets denoising with empirical Bayes thresholding (BWD) \cite{Johnstone2004}, nonlocal Bayesian image denoising (NLB) \cite{Lebrun2013}, anisotropic diffusion (AD) \cite{Weickert1998},  isotropic diffusion (ID) \cite{Perona1990, Bernardes2010}. We run GGD along with the above six denoising methods on five famous test images, namely, Barbara, boat, cameramen, clown, and mandrill, of size 150 $\times$ 150 that are downloaded from \cite{testImages}. We produce two versions of each test image by imposing two corruption levels $\epsilon$ = 40 and 80 using Eqn.~\eqref{eqn:noise}. We run GGD over all the 10 test images with arbitrary parameter values $\delta=20$, $\rho=7$, and $L=50$, and compute the reconstruction performance using the metrics RMSE (Def.~\ref{def:rmse}), PSNR (Def.~\ref{def:psnr}), and SSIM (Def.~\ref{def:ssim}) by treating the noise-free original image as the reference image  and the denoised image as the image of interest. 

\begin{table*}[!htp]
\caption{Comparison of the denoising performance, quantified using the metrics RMSE (the first row of each block), PSNR (the second row of each block), and SSIM (the third row of each block), of GGD and other six benchmark denoising methods, namely, sparse 3-D transform-domain collaborative filtering (BM3D), sparse and redundant representations over learned dictionaries (KSVD), Bayes wavelet denoising (BWD), nonlocal Bayesian image denoising (NLB), anisotropic diffusion (AD), and isotropic diffusion (ID). For each of the five test images, namely, cameraman, mandrill, Barbara, boat, and clown, of size 150 $\times$ 150, two noisy image instances are made by imposing them with two Gaussian noise distributions of mean zero and standard deviations ($\epsilon$'s) 40 and 80. The corruption levels of the input noisy images, computed using the same three metrics are denoted by $\mathcal{U}$. While GGD is executed with the parameter values $\delta=20$, $\rho=7$, and $L=50$, the other methods are executed using their recommended or default parameter values. For a given noisy image and a given performance metric, the colors red, blue, and green indicate the method performs the best, the second-best, and the third-best, respectively. We observed that the overall denoising performance of GGD is the best, that of BM3D is the second-best, and that of KSVD is the third-best.}\label{tab}
\begin{center}
\begin{tabular}{|p{.8 cm} | p{.7 cm} p{.7 cm} p{.7cm} p{.7cm} p{.7cm}| p{.7 cm} p{.7 cm} p{.7cm} p{.7cm} p{.7cm}|}
\hline
 \multirow{2}{*}{Meth.} & \multicolumn{5}{c|}{Images with $\epsilon = 40$} & \multicolumn{5}{c|}{Images with $\epsilon = 80$}\\
\hline
& Bar.  & Boa. & Cam. &  Clo. & Man. & Bar.  & Boa. & Cam. &  Clo. & Man. \\
 \hline
\multirow{3}{*}{$\mathcal{U}$} & 40.37 & 44.18 & 38.83 & 36.15 & 41.46 & 69.71 & 72.46 & 67.58 & 62.68 & 70.78 \\
& 16.01 & 15.23 & 16.35 & 16.97 & 15.78 & 11.27 & 10.93 & 11.53 & 12.19 & 11.13\\
 & .3444 &  .3067 & .2533 & .3676 & .3443 & .1627 & .1466 & .1359 & .1978 & .1628\\

\hline
\multirow{3}{*}{GGD} & \textcolor{red}{17.11} & \textcolor{red}{25.70} & \textcolor{teal}{18.18} & \textcolor{blue}{18.75} & \textcolor{red}{21.94} & \textcolor{red}{25.60} & \textcolor{red}{27.26} & 28.18 & \textcolor{red}{27.28} & \textcolor{red}{28.87}\\
& \textcolor{red}{23.47} & \textcolor{red}{20.87} & \textcolor{teal}{23.13} & \textcolor{blue}{22.94} & \textcolor{red}{21.56} &  \textcolor{red}{19.89} &  \textcolor{red}{19.42} & 19.13 & \textcolor{red}{19.41} & \textcolor{red}{18.85}\\
& \textcolor{red}{.7209} & \textcolor{blue}{.6070} & \textcolor{teal}{.6313} & \textcolor{red}{.6603} & \textcolor{red}{.5448} & \textcolor{red}{.5436} & \textcolor{blue}{.4578} & .4776 & \textcolor{red}{.4958} & \textcolor{red}{.3883} \\

 \hline
\multirow{3}{*}{BM3D} & \textcolor{blue}{18.79} & \textcolor{blue}{25.87} &  \textcolor{red}{14.36} &  \textcolor{red}{18.01} & \textcolor{teal}{22.33} & 30.42 & \textcolor{blue}{32.57} & \textcolor{red}{23.89} & \textcolor{teal}{28.43} & 29.22\\
 &  \textcolor{blue}{22.65} & \textcolor{blue}{19.87} & \textcolor{red}{24.99} & \textcolor{red}{23.02} & \textcolor{teal}{21.15} & 18.47 & \textcolor{blue}{17.88} & \textcolor{red}{20.57} & \textcolor{teal}{19.06} &  18.78 \\
 &  \textcolor{blue}{.7195}  &  \textcolor{red}{.6365}  &  \textcolor{red}{.7531} &  \textcolor{blue}{.6594}  &  .4852 & .5043 & \textcolor{red}{.4829} &  \textcolor{red}{.6294} &  \textcolor{teal}{.4847} & .3370\\

 \hline
\multirow{3}{*}{KSVD} & \textcolor{teal}{18.88} & \textcolor{teal}{26.06} & \textcolor{blue}{17.09} & \textcolor{teal}{18.83} & \textcolor{blue}{21.97} & \textcolor{blue}{27.96} & \textcolor{teal}{33.17} & \textcolor{teal}{25.91} & \textcolor{blue}{27.93} &  29.54\\
&  \textcolor{teal}{22.61} & \textcolor{teal}{19.81} & \textcolor{blue}{23.48} & \textcolor{teal}{22.63} & \textcolor{blue}{21.29} & \textcolor{blue}{19.20} & \textcolor{teal}{17.71} & \textcolor{teal}{19.86} & \textcolor{blue}{19.21} & 18.72\\
&   \textcolor{teal}{.7008}  &  \textcolor{teal}{.6053}  &  .6088 & \textcolor{teal}{.6348} & \textcolor{blue}{.5347} & \textcolor{blue}{.5251} & .4299 & .4131 & \textcolor{blue}{.4789} &  .3336\\

\hline
\multirow{3}{*}{BWD} & 21.91 & 28.53 & 19.59 & 21.64 & 24.67 & 30.22 & 35.09 & 27.92 & 31.28 & 30.05\\
&   21.32 & 19.02 & 22.29 & 21.43 & 20.29 & 18.52 & 17.23 & 19.21 & 18.23 & 18.57 \\
 &   .6373  &  .5691  &  \textcolor{blue}{.6438}  &  .5871  &  .4041 & .4987 & .4491 & \textcolor{blue}{.5287} & .4381 & .3292 \\

 \hline
\multirow{3}{*}{NLB} & 22.62 & 29.56 & 21.07 & 20.41 & 25.74 & 36.30 & 41.12 &  34.16 & 33.38 & 38.00 \\
 &  21.04 & 18.72 & 21.66 & 21.94 & 19.92 & 16.93 & 15.85 & 17.46 & 17.66 & 16.53\\
 &   .5647  &  .4696  &  .4032  &  .5527  &  .5005 & .3449 & .2822 & .2533 & .3630 & .3007\\
 
\hline
\multirow{3}{*}{AD} & 21.67 & 27.91 & 19.72 & 22.21 & 22.36 & \textcolor{teal}{29.08} & 33.65 & 26.68 &  30.12 & \textcolor{blue}{28.90}\\
 &  21.42 & 19.22 &  22.23 & 21.19 & 21.14 & \textcolor{teal}{18.86} & 17.59 & 19.61 & 18.55 & \textcolor{blue}{18.84}\\
 &   .6535 &   .5807 &   .5859 &   .5895 &   \textcolor{teal}{.5217} & .4967 & .4278 & .4064 & .4470 & \textcolor{teal}{.3849}\\

\hline
\multirow{3}{*}{ID} & 24.27 & 29.53 & 19.63 & 23.54 & 24.47  & 29.87 & 33.95 & \textcolor{blue}{25.88} & 30.37 & \textcolor{teal}{29.03} \\
  & 20.43 & 18.73 & 22.27 & 20.70 & 20.36 & 18.63 & 17.51 & \textcolor{blue}{19.87} & 18.48 & \textcolor{teal}{18.80} \\
&  .6204  &  .5698  &  .6884  &  .5738  &  .4222 & \textcolor{teal}{.5189} & \textcolor{teal}{.4538} & \textcolor{teal}{.5134} & .4770 & \textcolor{blue}{.3856}\\
\hline
\end{tabular}
\end{center}
\end{table*}

We set the parameters of the other six denoising methods to the recommended values in their literature or to the default values if nothing is recommended. The parameters denoising strength, patch size, sliding step size, the maximum number of similar blocks, radius for search block matching, step between two search locations, 2D thresholding, 3D thresholding, and threshold for the block-distance of BM3D are set to $\epsilon$ (the standard deviation of the imposed noise in GGD), 12, 4, 16, 39, 1, 2, 2.8, and 3000, respectively. Block size, dictionary size, number of training iterations, Lagrangian multiplier, noise gain, and number of non-zero coefficients of KSVD are set to 64, 244, 10, 30/$\epsilon$, 1.55, and 2, respectively. The wavelet decomposition level of BWD is set to 3. Patch size, number of similar patches, search window size, truncated rank, collaborative filtering coefficient, and the minimum threshold for similar patches of NLB are set to 8, 7, 40, 1.05, 1, and 4, respectively. The standard deviation of the edge-stopping function, amount of diffusion, and maximum iterations of AD to 1.5, 2, and 300, respectively. 
The standard deviation of the edge-stopping function, amount of diffusion, and maximum iterations of ID are set to 5, 5, and 1000, respectively. We run the above six methods over these 10 test images with the above parameter values and compute the reconstruction errors using the same three performance metrics RMSE, PSNR, and SSIM by treating the noise-free original image as the reference image  and the denoised image as the image of interest. We also compute the level of corruption of each test image using the same three performance metrics error metric and use them to compare the denoising results between methods.

\begin{figure*}[htp]
	\begin{center}	
	\includegraphics[width=1\textwidth]{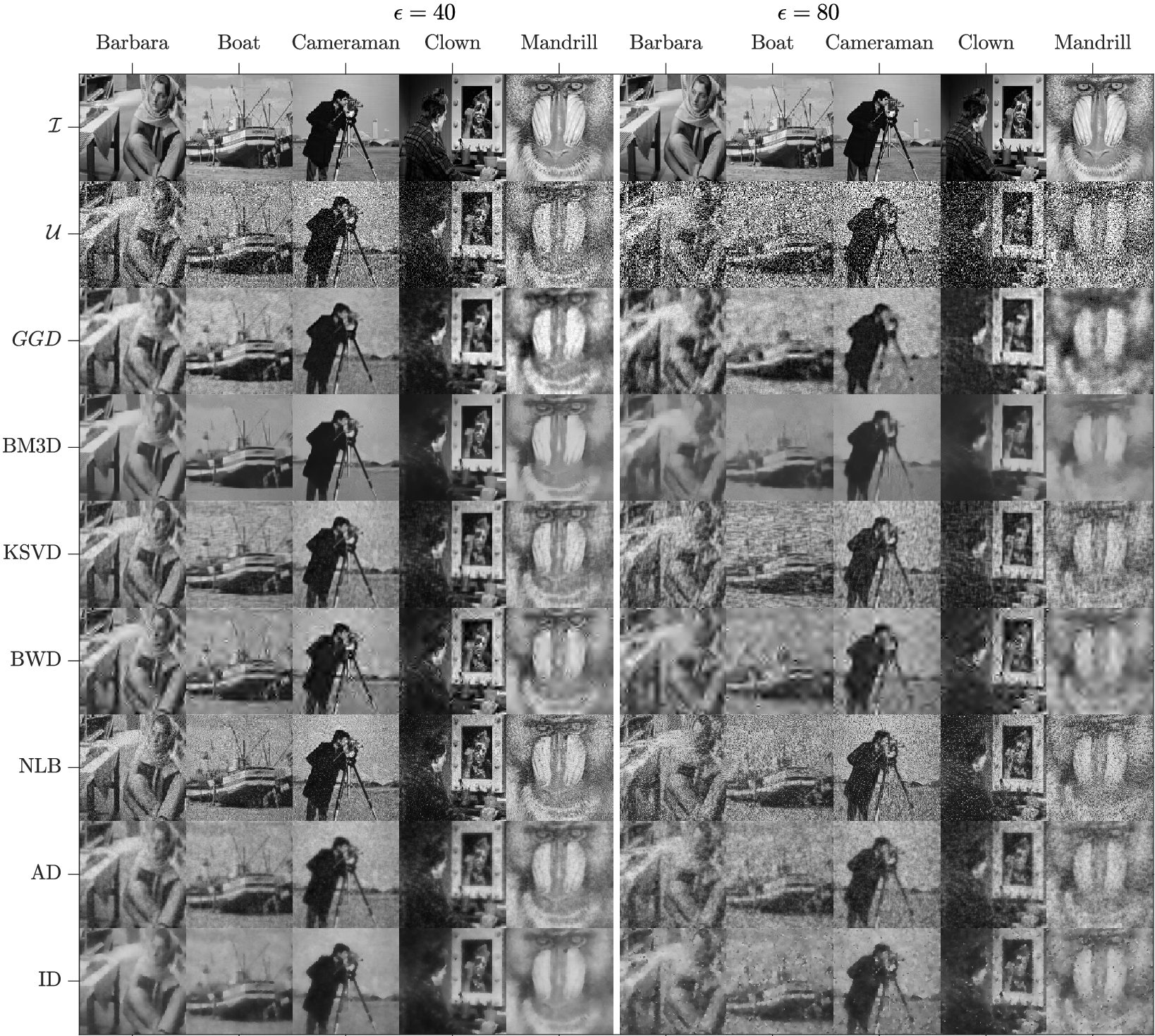}
	\end{center}
	\caption{Visual comparison of the quality of the images denoised by GGD with that of the six other benchmark denoising methods, namely, sparse 3-D transform-domain collaborative filtering (BM3D), sparse and redundant representations over learned dictionaries (KSVD), Bayes wavelet denoising (BWD), nonlocal Bayesian image denoising (NLB), anisotropic diffusion (AD), and isotropic diffusion (ID). Each of the five test images, namely, cameraman, mandrill, Barbara, boat, and clown, of size 150 $\times$ 150, is imposed with two Gaussian noise distributions with a mean of 0 and standard deviations ($\epsilon$'s) of 40 and 80. Here, the original noise-free images are denoted by $\mathcal{I}$'s and their noisy versions are denoted by $\mathcal{U}$'s (see supplementary materials for the enlarged images). While GGD is executed with the parameter values $\delta=20$, $\rho=7$, and $L=50$, the other methods are executed with their recommended or default parameter values. Here, we observe that GGD retains texture and cartoon in the denoised images most of the time than that of the other six methods.}
	\label{fig:diffMethod}
\end{figure*}

Table~\ref{tab} presents the corruption levels, computed using the metrics RMSE, PSNR, and SSIM, of the 10 noisy test images, see the row of $\mathcal{U}$, and the reconstruction errors, computed using the same metrics, of the images denoised by the eight methods including GGD. Therein, for a given noisy image and a given performance metric, the colors red, blue, and green represent the denoising method performing  the best, the second-best, and the third-best performance, respectively. We observe that GGD obtains 20 reds, 4 blues, and 3 greens; BM3D obtains 10 reds, 8 blues, and 5 greens; and KSVD obtains 11 blues and 13 greens. This observation evidences that the overall denoising performance of GGD is the best, that of BM3D is the second-best, and that of KSVD is the third-best. Fig.~\ref{fig:diffMethod} visualizes the quality of the denoising of the 10 test images where we observe that GGD performs better most of the time than all of the other six methods in terms of preserving both the texture and cartoon of the original images. 

\section{Conclusion}\label{sec:conclusion}
In this paper, we introduced a novel image denoising method that utilizes eigenvectors of the Gramian matrix of the graph geodesics evaluated on the nosy image's patch-space. Specifically, we partitioned a given noisy image into overlapping square-shaped patches with a known length, say $\rho$, where each patch is a point in a high-dimensional space of $\rho^2$-dimensions. A low-dimensional manifold underlies this high-dimensional data cloud of the patch-set characterizes features of the noisy image. This manifold was formulated using the eigenvectors corresponding to the biggest eigenvalues of the Gramian matrix of the graph geodesic distances measured on the nosy image's patch-space. Specifically, we produced a graph structure by treating the high-dimensional patches as vertices and by joining nearest neighbor vertices to each vertex with edges of Euclidean distances between them. The geodesic distance between two patches is estimated as the graph shortest path between them. Then, we transformed the geodesic distance matrix into its Gramian matrix. The prominent eigenvectors of the Gramian matrix were used to produce the noise-free patches and these noise-free patches were merged to generate the denoised image. The reason for the adoption of geodesic distance over Euclidean distance as a proximity for the manifold distance is that the geodesic distance is nonlinear whereas the Euclidean distance is linear. Nonlinear proximity of a manifold mimics the manifold closely so that it helps capturing the true geometry of the manifold underlying the image patch-space; thus, it ensures better quality of the denoised images that retains essential image features.

We observed that when the noise contamination increases the patch size should be increased to obtain better denoising performance. This is because, if the image possesses a high noise, a big patch size helps smooth the image more; however, too big patch size might smooth the image too much. The neighborhood size doesn't have a significant influence on the denoising performance except for the case where the patch size is small. The denoising performance decreases slightly when the neighborhood size increases, especially for significantly small patch sizes. The reason for that is big neighborhood sizes add more edges into the graph structure and that causes short-circuiting of the network when the geodesics are approximated. These  underestimated geodesic distances on the manifold lead to less denoising performance.

Patch size, neighborhood size, and eigenvector threshold are the only three user input parameters in GGD whereas most of the similar patch-based non-local denoising methods, such as BM3D, KSVD, and NLB, are well known to have many user input parameters. We have seen that the parameter neighborhood size  of GGD is less influential for the performance that the user may set to a common value for all the experiments. For an image with high noise, our recommendation is to use a slightly bigger neighborhood size (e.g.~20) and a slightly bigger patch size (e.g.~9), and a slightly smaller eigenvector threshold (e.g.~10) in contrast to that for an image with low noise. Methods with a variety of highly influential parameters are inconvenient to use since either the user has to set a trial and error procedure to search on the entire parameter domain to find the best values for the parameters or has to make careful guesses for their values. While making a personal parameter guess is highly subjective and depends on personal experience, a trial and error procedure to search the entire parameter domain consumes a significant time and requires more computational power. 

We ran the same experiment for several realizations with different random seeds in the probability distribution that we sample the noise if the experiment is to test the influence of noise contamination on denoising performance. The errorbars in the figures represent the standard deviation of the reconstruction performance across the realizations. For each performance metric, we observed that the errorbars of the plots corresponding to the denoised images associated with small eigenvector thresholds  are smaller than that of the initial noisy images. Smaller errorbars infer the stability of the denoising method across realizations. Due to the fact that small eigenvector thresholds remove more noise from the denoising, that will also reduce the influence of the choice of the random seed on the performance. This independent aspect across different noise samples is essential for denoising methods since the noise in images is natural so the user doesn't know how the noise is sampled from the underlying probability distribution. 

We validated the performance of GGD against six benchmark denoising algorithms, namely, sparse 3-D transform-domain collaborative filtering (BM3D), sparse and redundant representations over learned dictionaries (KSVD), Bayes wavelet denoising (BWD), nonlocal Bayesian image denoising, anisotropic diffusion, Bayesian estimation denoising, and isotropic diffusion. GGD preserves both the texture, containing edges and corners, and the cartoon, containing piece-wise smooth parts, of the original images to high accuracy than that of the other six methods. Specifically, GGD performs better than the renowned methods such as BM3D and KSVD, and than the commercially implemented method BWD that is available in MATLAB. For all the methods, we observe that the denoising performance associated with the test image boat is the lowest than that of the other four test images due to the fact that this image possesses more texture than that of the other four test images.

GGD uses eigenvalue decomposition to generate eigenvectors of the Gramian matrix where this matrix is $n^2 \times n^2$ for an image of $n\times n$ pixels. Since the computational complexity of the eigenvalue decomposition of a matrix of size $m\times m$ is $m^3$ \cite{Lee2009}, the computational complexity of the eigenvectors of this Gramian matrix is significantly expensive as $n^6$. Thus, to overcome this issue, computation of only the required fewer number of eigenvectors of this Gramian matrix is worthwhile. GGD always performs better denoising with small eigenvector thresholds independent of the nature of the input image or the other parameters in-use. Since we currently compute the entire eigenvector spectrum and use only the prominent ones of them in GGD, in the future, we will improve GGD by incorporating eigenvector estimation strategies that are available in the literature of \emph{Bigdata}. For that, we are planning to replace the regular eigenvalue decomposition routing with multiple eigenvalue approximation strategies such as, 1) inverse-free preconditioned \emph{Krylov subspace method} called \emph{Lanczos algorithm} \cite{Liang2014, Saad2011}; 2) \emph{random sampling} method that trains a neural network of rows of the matrix \cite{Kobayashi2001}; 3) \emph{Monte Carlo} approach that iteratively makes approximations \cite{Liang2014, Friedland2006}. This future work will reduce the computational complexity of GGD significantly so that the method can be brought into a stage where it can be integrated into real-time image denoising applications.

\section{Compliance with Ethical Standards}
The authors have no relevant financial or non-financial interests to disclose.

\section{Data Availability Statements}
The datasets generated during and/or analyzed during the current study are available from the corresponding author upon reasonable request.


\end{document}


\maketitle\vspace{-1cm}
\rule{17cm}{0.4pt}\\

Supplementary materials in this document are the enlarged test images, of the experiments presented in the journal article ``Image Denoising Using the Geodesics' Gramian of the Manifold Underlying Patch-Space'', for better visualization. Here, we denote a Gaussian distribution of mean zero and standard deviation $\epsilon$ by $\mathcal{N}(0,\epsilon^2)$.

\begin{figure}[htp]
	\begin{center}
	\includegraphics[width=1\textwidth]{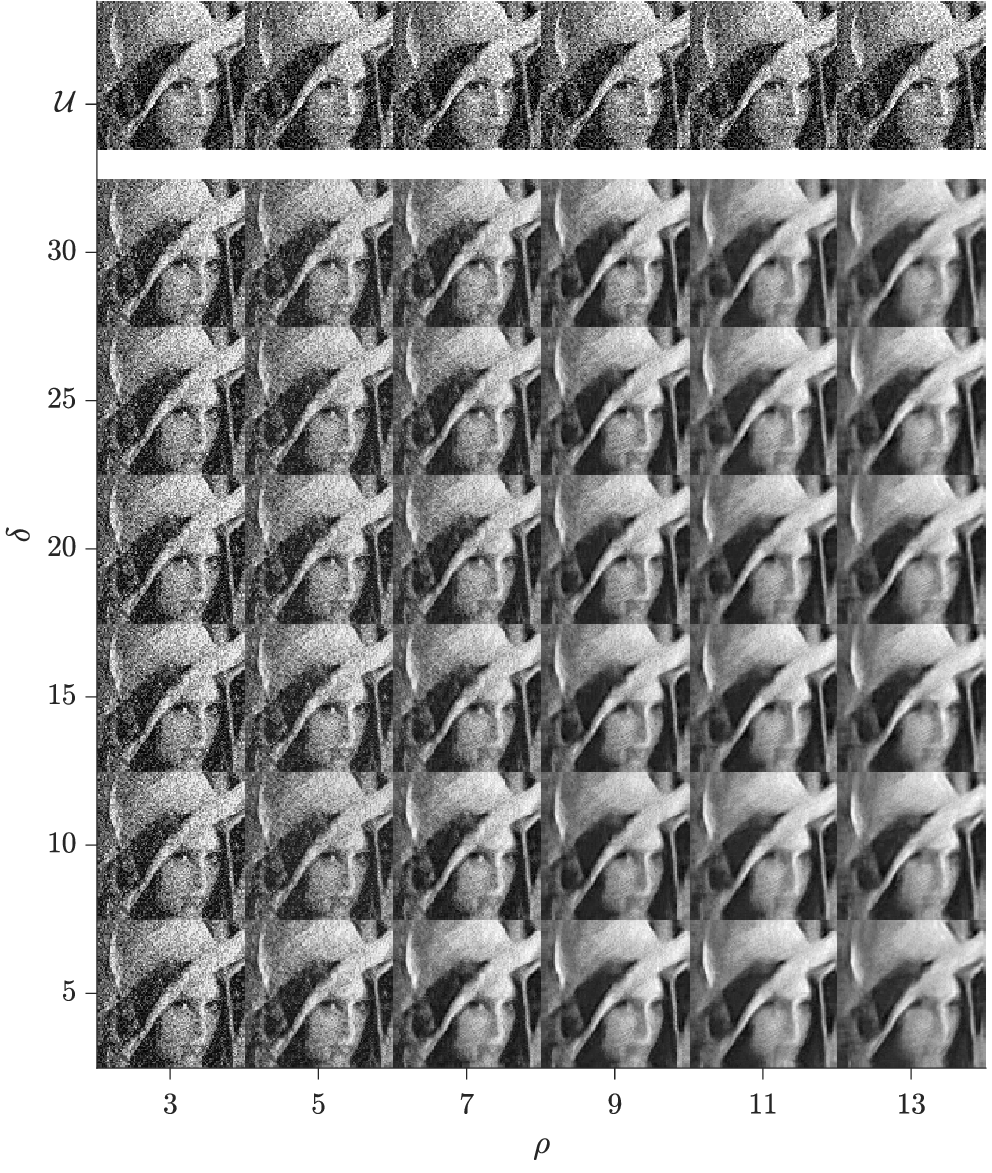}
	\end{center}
	\caption{Fig.~2(j) of the journal articles where the test image is imposed with Gaussian noise sampled from the distributions $\mathcal{N}(0,40^2)$. Here, $\mathcal{U}$ denotes the original noisy image, $\rho$ denotes patch size, and $\delta$ denotes neighborhood size. While the first row is the same noisy image that is repeated through the columns, the rest of them are the images denoised by GGD with different combinations of $\rho$ and $\delta$.}
	\label{fig:rhovsk1}
\end{figure}

\begin{figure}[htp]
	\begin{center}
	\includegraphics[width=1\textwidth]{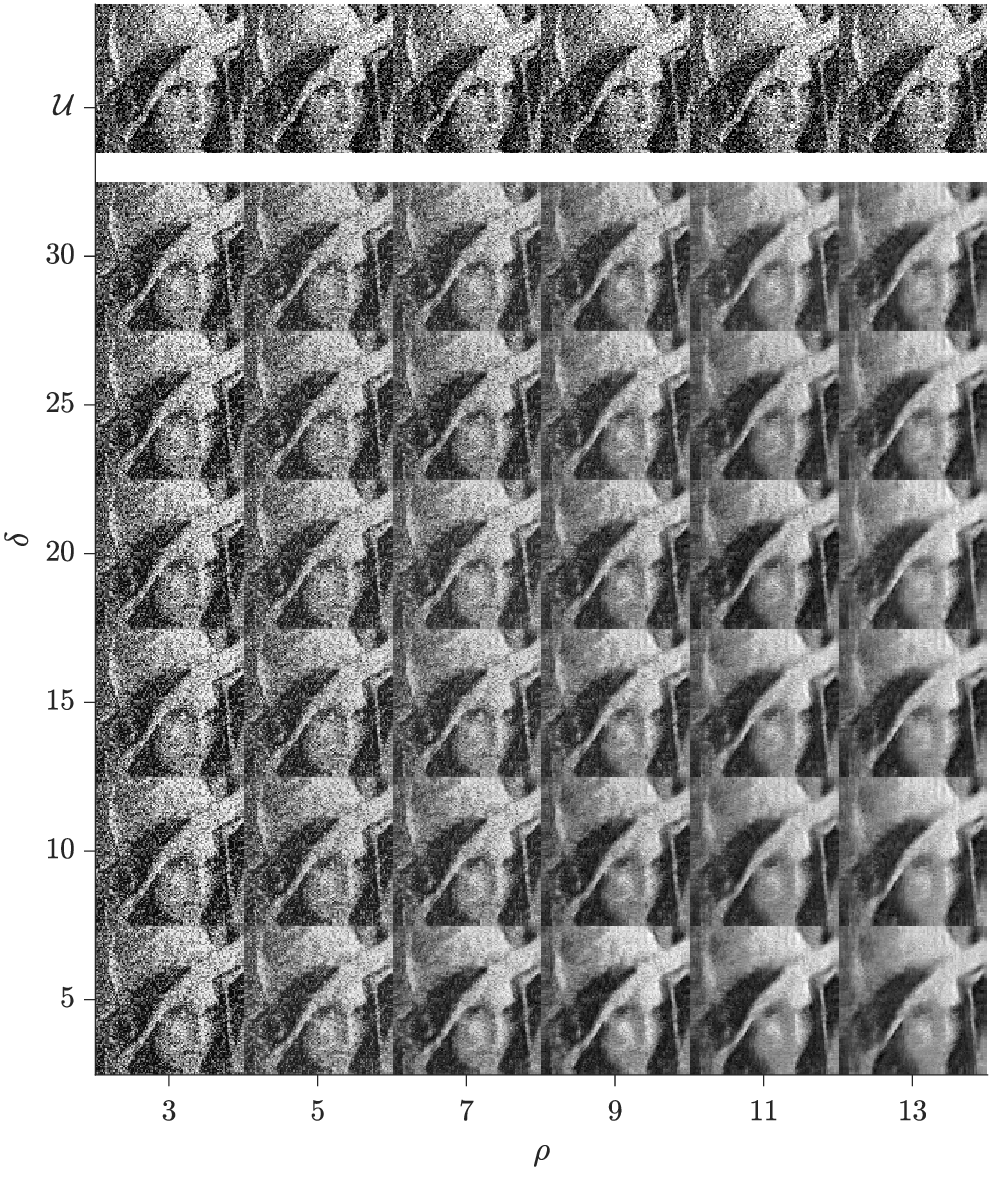}
	\end{center}
	\caption{Fig.~2(k) of the journal articles where the test image is imposed with Gaussian noise sampled from the distributions $\mathcal{N}(0,60^2)$. Here, $\mathcal{U}$ denotes the original noisy image, $\rho$ denotes patch size, and $\delta$ denotes neighborhood size. While the first row is the same noisy image that is repeated through the columns, the rest of them are the images denoised by GGD with different combinations of $\rho$ and $\delta$.}
	\label{fig:rhovsk2}
\end{figure}

\begin{figure}[htp]
	\begin{center}
	\includegraphics[width=1\textwidth]{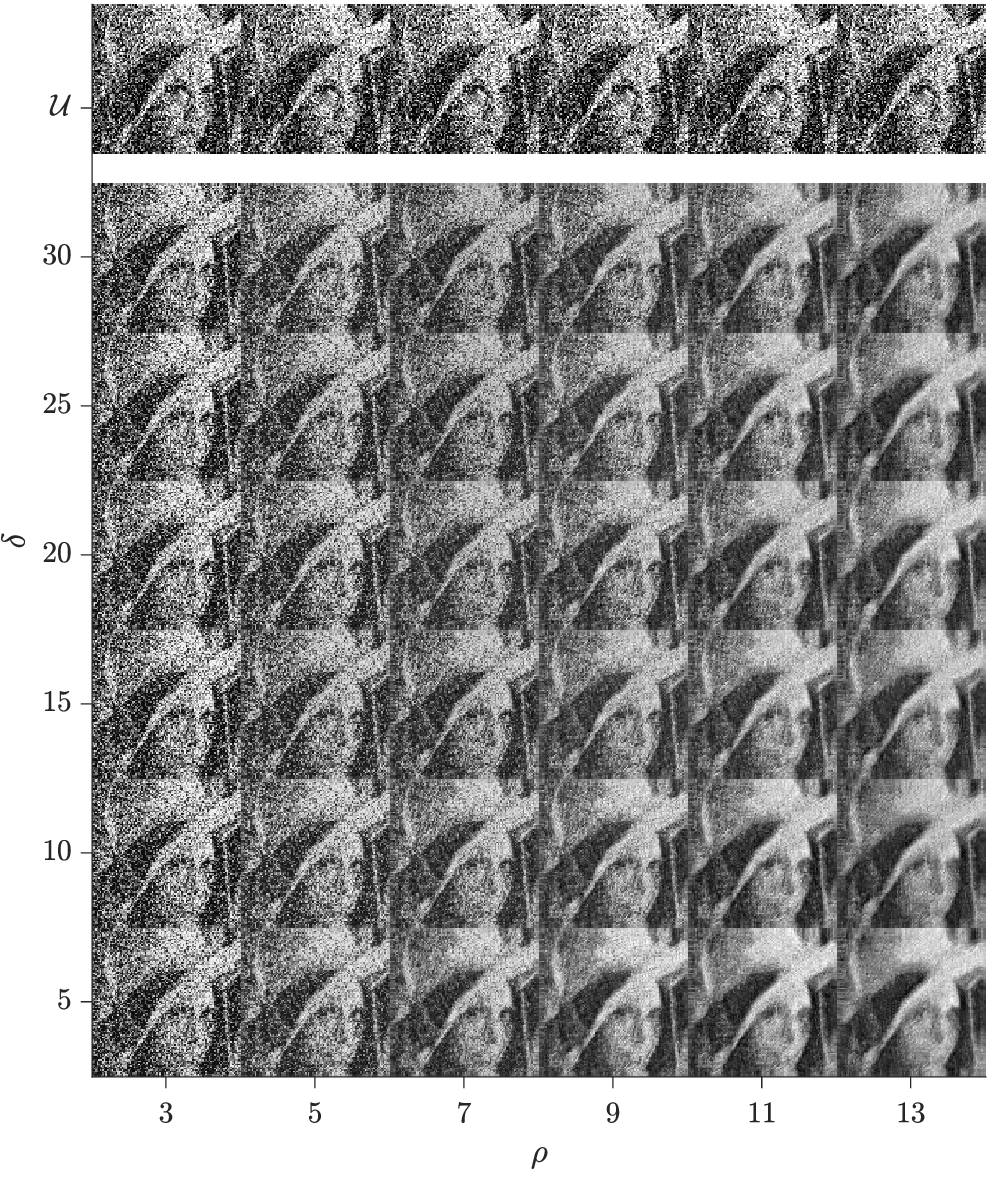}
	\end{center}
	\caption{Fig.~2(l) of the journal articles where the test image is imposed with Gaussian noise sampled from the distributions $\mathcal{N}(0,80^2)$. Here, $\mathcal{U}$ denotes the original noisy image, $\rho$ denotes patch size, and $\delta$ denotes neighborhood size. While the first row is the same noisy image that is repeated through the columns, the rest of them are the images denoised by GGD with different combinations of $\rho$ and $\delta$.}
	\label{fig:rhovsk3}
\end{figure}

\begin{figure}[htp]
	\begin{center}
	\includegraphics[width=.5\textwidth]{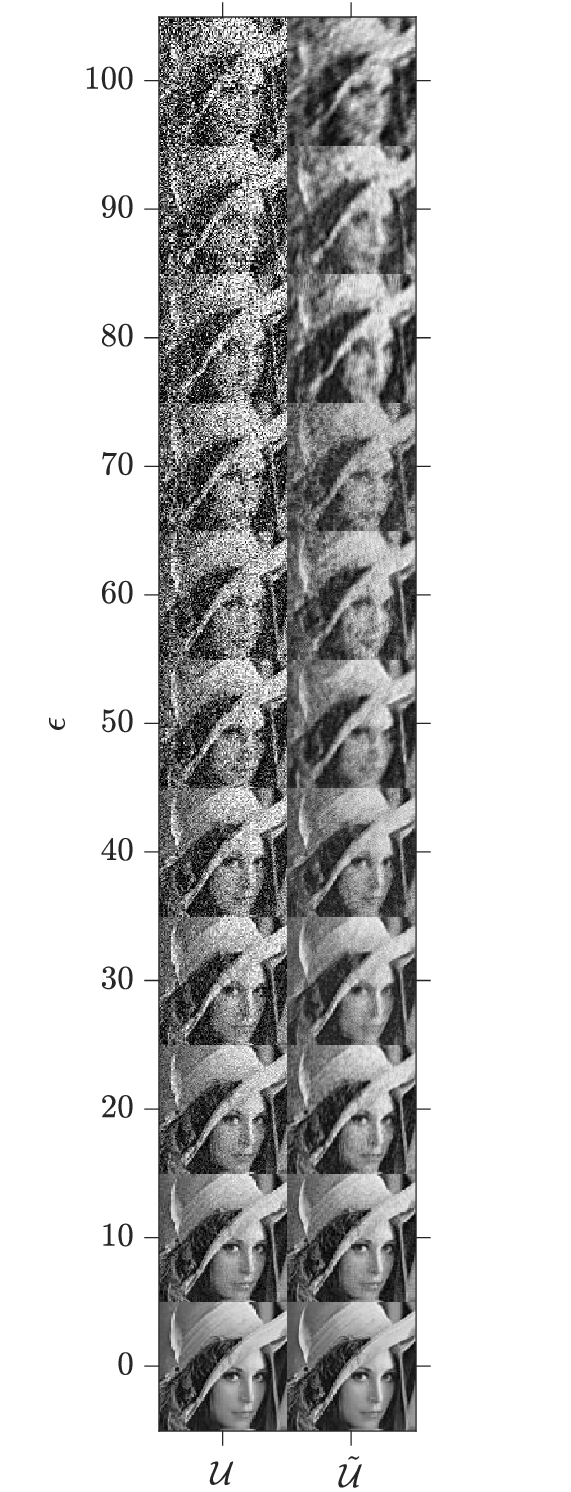}
	\end{center}
	\caption{Fig.~3(d) of the journal article  where the test image is imposed with different levels of Gaussian noise sampled from the distributions $\mathcal{N}(0,\epsilon^2)$, where $\epsilon=0,10,\dots,100$. Here, $\mathcal{U}$ denotes original noisy images and $\tilde{\mathcal{U}}$ denotes the images denoised by GGD.}
	\label{fig:3d}
\end{figure}

\begin{figure}[htp]
	\begin{center}
	\includegraphics[width=.8\textwidth]{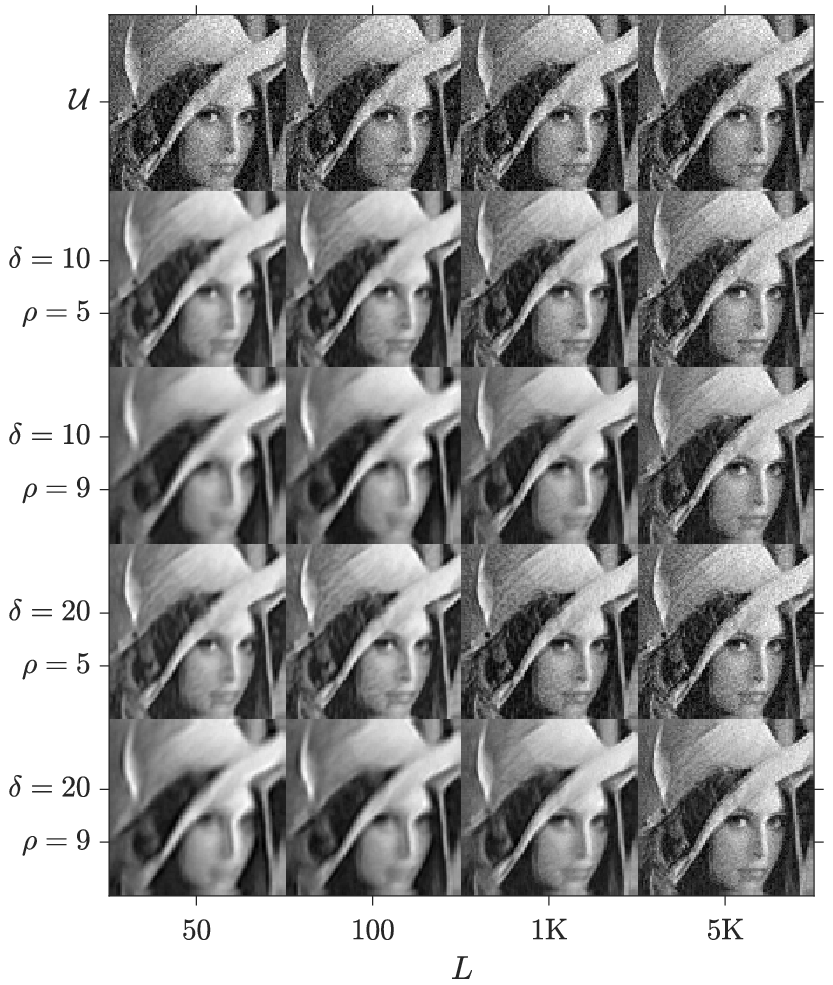}
	\end{center}
	\caption{Fig.~4(j) of the journal article where the test image is imposed with Gaussian noise sampled from the distribution $\mathcal{N}(0,40^2)$. Here, $\mathcal{U}$ denotes original noisy images, $\rho$ denotes patch size, $\delta$ denotes neighborhood size, and $L$ denotes the eigenvector threshold. While the first row presents the same noisy image repeatedly throughout the columns, the rest of them are the images denoised by GGD with different combinations of $\rho$, $\delta$, and $L$.}
	\label{fig:4j}
\end{figure}

\begin{figure}[htp]
	\begin{center}
	\includegraphics[width=.8\textwidth]{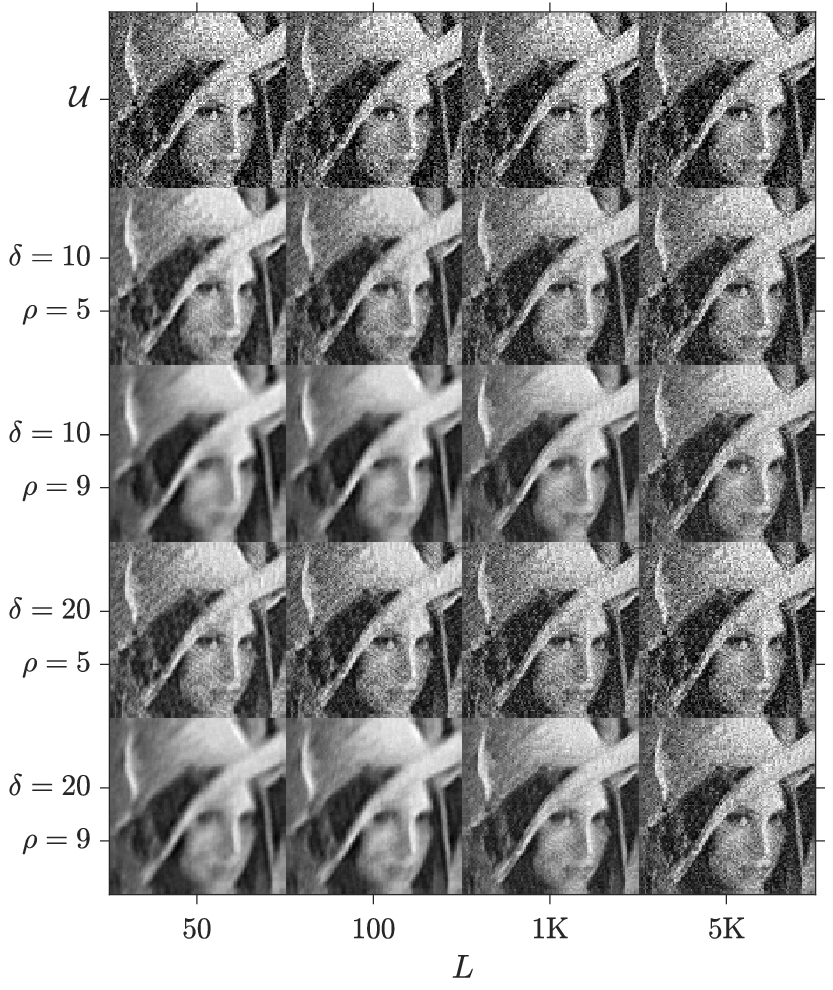}
	\end{center}
	\caption{Fig.~4(j) of the journal article where the test image is imposed with Gaussian noise sampled from the distribution $\mathcal{N}(0,60^2)$. Here, $\mathcal{U}$ denotes original noisy images, $\rho$ denotes patch size, $\delta$ denotes neighborhood size, and $L$ denotes the eigenvector threshold. While the first row presents the same noisy image repeatedly throughout the columns, the rest of them are the images denoised by GGD with different combinations of $\rho$, $\delta$, and $L$.}
	\label{fig:4k}
\end{figure}

\begin{figure}[htp]
	\begin{center}
	\includegraphics[width=.8\textwidth]{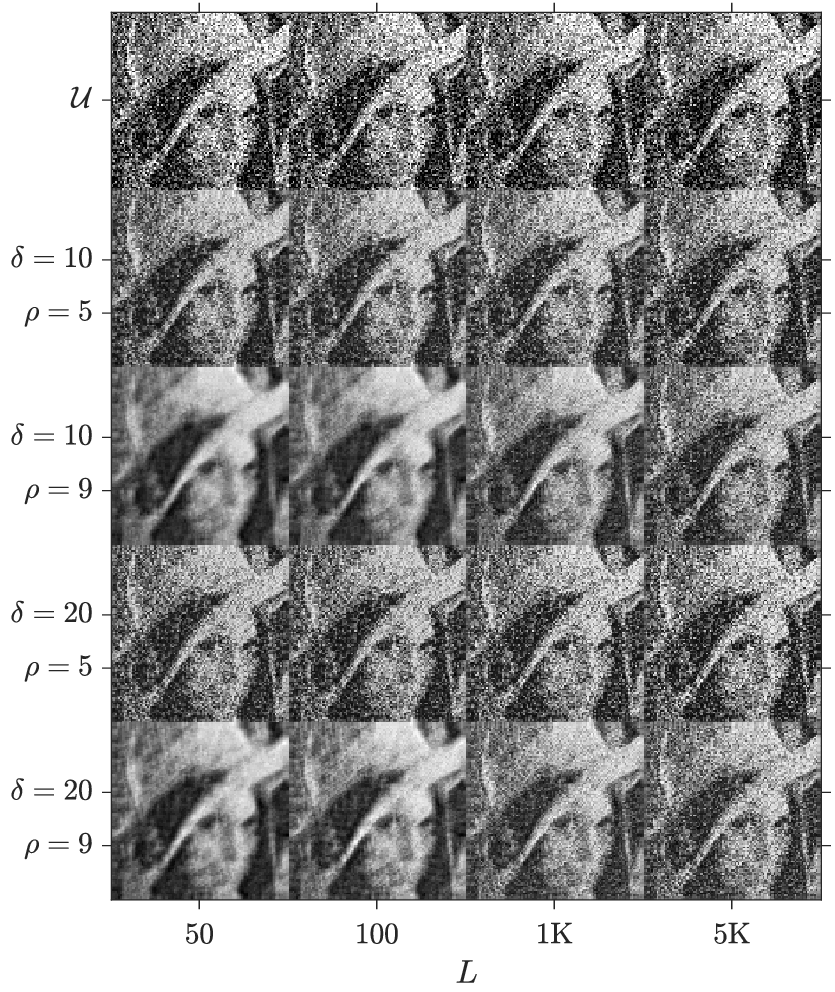}
	\end{center}
	\caption{Fig.~4(k) of the journal article where the test image is imposed with Gaussian noise sampled from the distribution $\mathcal{N}(0,80^2)$. Here, $\mathcal{U}$ denotes original noisy images, $\rho$ denotes patch size, $\delta$ denotes neighborhood size, and $L$ denotes the eigenvector threshold. While the first row presents the same noisy image repeatedly throughout the columns, the rest of them are the images denoised by GGD with different combinations of $\rho$, $\delta$, and $L$.}
	\label{fig:4l}
\end{figure}

\begin{figure}[htp]
	\begin{center}
	\includegraphics[width=.8\textwidth]{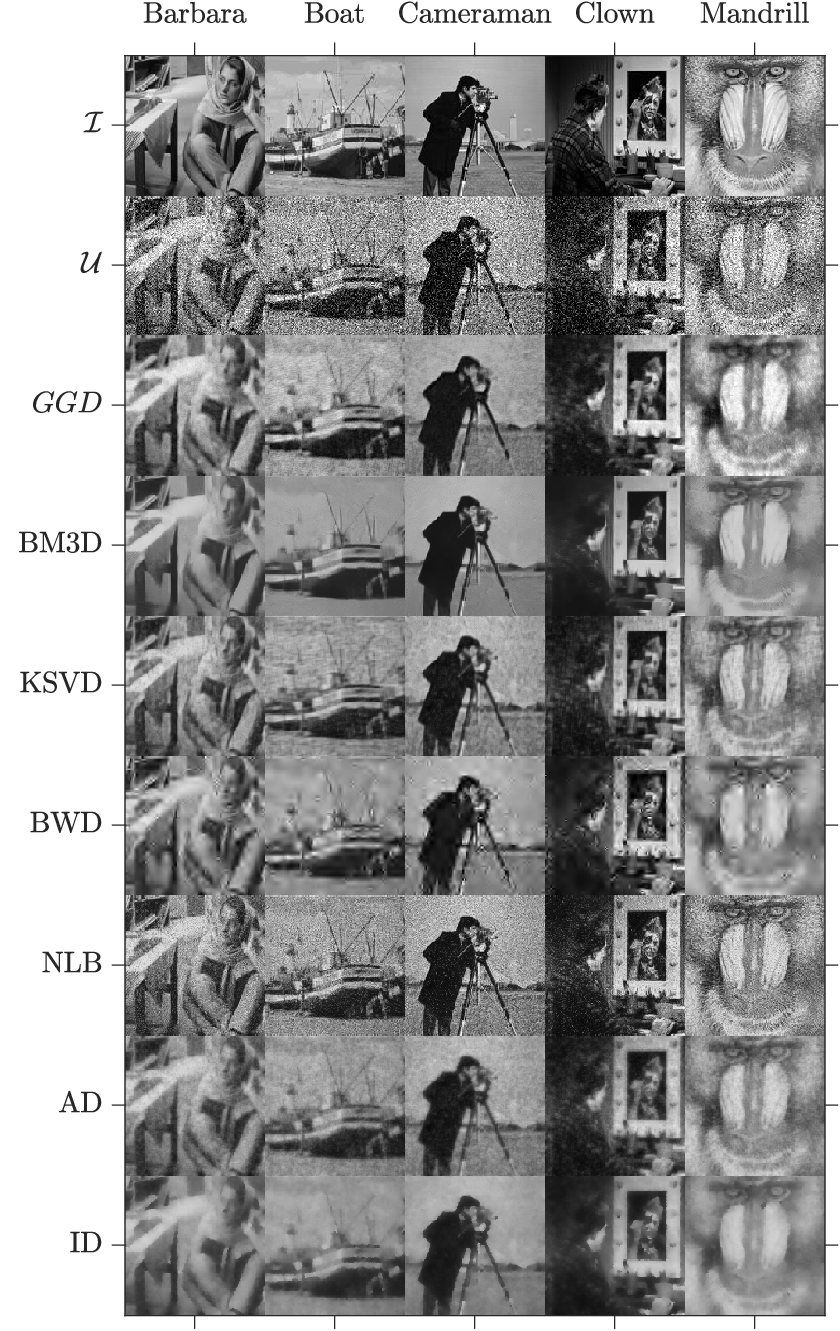}
	\end{center}
	\caption{Right-half of Fig.~5 of the journal article where five test images are imposed with Gaussian noise sampled from the distribution $\mathcal{N}(0,40^2)$. Here, $\mathcal{I}$ denotes the original noisy-free images and $\mathcal{U}$ denotes the original noisy images. While GGD abbreviates Geodesic Gramian Denoising and represents its denoising, the rest of the abbreviations stand for the other six benchmark denoising methods and they represent their denoising.}
	\label{fig:5a}
\end{figure}

\begin{figure}[htp]
	\begin{center}
	\includegraphics[width=.8\textwidth]{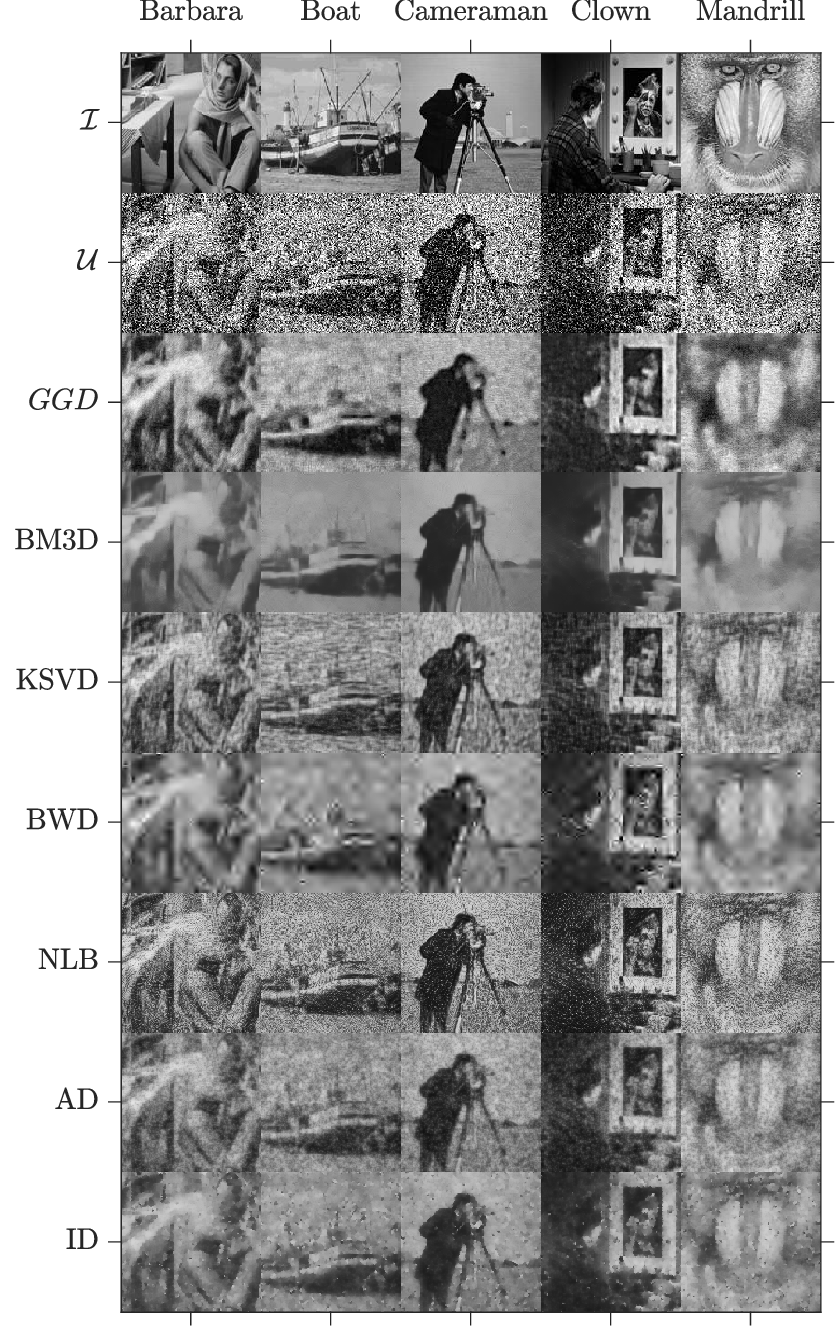}
	\end{center}
	\caption{Left-half of Fig.~5 of the journal article where five test images are imposed with Gaussian noise sampled from the distribution $\mathcal{N}(0,80^2)$. Here, $\mathcal{I}$ denotes the original noisy-free images and $\mathcal{U}$ denotes the original noisy images. While GGD abbreviates Geodesic Gramian Denoising and represents its denoising, the rest of the abbreviations stand for the other six benchmark denoising methods and they represent their denoising.}
	\label{fig:5b}
\end{figure}